\PassOptionsToPackage{greek.polutoniko,english}{babel}
\documentclass[letterpaper,USenglish]{lipics-v2021}

\nolinenumbers
\hideLIPIcs

\AtBeginDocument{%
  \providecommand\BibTeX{{%
    \normalfont B\kern-0.5em{\scshape i\kern-0.25em b}\kern-0.8em\TeX}}}

\AtBeginDocument{%
    \setlength\intextsep{2pt}%
    \setlength\textfloatsep{2pt}%
    \setlength\floatsep{2pt}}

\hypersetup{hidelinks=false,colorlinks=true,linkcolor=blue,urlcolor=blue,citecolor=blue,anchorcolor=blue}

\usepackage{url} 
\usepackage{tabularx} 
\usepackage{booktabs} 
\usepackage{subcaption} 
\usepackage{placeins}       
\usepackage{stfloats}          
\usepackage{textgreek} 
\usepackage{pdflscape}          

\usepackage[dvipsnames]{xcolor}

\usepackage[table]{xcolor}
\definecolor{lightgray}{gray}{0.94}
\let\oldtabular\tabular
\let\endoldtabular\endtabular
\renewenvironment{tabular}{\rowcolors{2}{white}{lightgray}\oldtabular}{\endoldtabular}

\usepackage[maxfloats=150]{morefloats}                                               
\usepackage{microtype}                                                               
\usepackage[utf8]{inputenc}   

\newcommand\Invisible[1]{                                                            
  \marginpar{\color{white}{\fontsize{.5}{.5}\selectfont #1 }}                        
}

\newcommand\InvisibleA[1]{}

\newcommand\Texticle{\vspace{0.25 \baselineskip}\noindent\textbf{$\blacktriangleright$} }

\newcommand{\Exclude}[1]{}

\newcommand\BoldSection[1]{\vspace{0.40 \baselineskip} \noindent \textbf{#1} \noindent}
\definecolor{Gray95}{gray}{0.95}
\definecolor{forestgreen}{rgb}{0.13, 0.55, 0.13}

\newcommand{\remove}[1] {}

\newcommand{\AtFoot}[1]{\let\thefootnote\relax\footnotetext{{#1}}}

\usepackage{rotating} 
\newcommand{\rot}[1]{\makebox{\rotatebox{90}{#1}}}%

\usepackage{calc}       

\newcommand{\blu}[1]{\colorbox{blue!40}{#1}}
\newcommand{\red}[1]{\colorbox{red!40}{#1}}  

\makeatletter
\newcommand{\setword}[2]{%
  \phantomsection
  #1\def\@currentlabel{\unexpanded{#1}}\label{#2}%
}
\makeatother

\makeatletter
\newcommand{\namelabel}[1]{%
  \phantomsection
  \renewcommand{\@currentlabel}{#1}
  \label{#1}
}
\makeatother

\usepackage{microtype}   
\usepackage{listings} 
\usepackage{graphicx} 
\usepackage{bold-extra} 
 
\usepackage{ifxetex}
\ifxetex
   \usepackage{fontspec}
\else
\fi

\makeatletter
\renewcommand*\verbatim@nolig@list{}
\makeatother

\usepackage[T1]{fontenc}
\usepackage[newfloat=true,frozencache=true]{minted}
\usepackage{changepage} 

\usepackage[iso,american,cleanlook]{isodate}                                         

\usepackage[export]{adjustbox} 

\usepackage{fancyhdr}                                                             
\makeatletter\let\ps@plain\ps@fancy\makeatother
\makeatletter
\lst@Key{countblanklines}{true}[t]%
    {\lstKV@SetIf{#1}\lst@ifcountblanklines}

\lst@AddToHook{OnEmptyLine}{%
    \lst@ifnumberblanklines\else%
       \lst@ifcountblanklines\else%
         \advance\c@lstnumber-\@ne\relax%
       \fi%
    \fi}
\makeatother

\lstdefinestyle{numbers}
{numbers=left, stepnumber=1, numberstyle=\tiny, numbersep=10pt}
\lstdefinestyle{nonumbers}
{numbers=none}

\usepackage{circledtext}  
\usepackage{circlesteps} 
\usepackage{tikz} 

\usepackage{paralist} 

\usepackage[inline]{enumitem}

\circledtextset{boxcolor=cyan,boxtype=o,charf=\huge,boxfill = blue!60}

\pgfkeys{/csteps/inner color=yellow}
\pgfkeys{/csteps/outer color=red}
\pgfkeys{/csteps/fill color=black}

\usepackage{amsmath}
\usetikzlibrary{arrows}
\usetikzlibrary{decorations.pathreplacing}
\usetikzlibrary{calligraphy} 

\usepackage{orcidlink} 

\usepackage[ampersand]{easylist}
\usepackage[margin=15pt,font=small,labelfont={bf,sf}]{caption}

\usepackage{fancyhdr}                                                             

\title{Hapax Locks} 
\subtitle{Value-Based Mutual Exclusion} 
\author{Dave Dice}{\strut}{david.dice@gmail.com}{https://orcid.org/0000-0001-9164-7747}{}
\author{Alex Kogan}{Oracle Labs}{alex.kogan@oracle.com}{https://orcid.org/0000-0002-4419-4340}{}                                  
\authorrunning{Dice and Kogan} 
\titlerunning{Hapax Locks} 
\Copyright{Oracle and/or its affiliates}
\acknowledgements{We thank Peter Buhr at the University of Waterloo for access to ARMv8 and AMD systems.  
We also thank the anonymous PPoPP reviewers for their helpful comments.} 
\relatedversion{This is an expanded version of the conference paper appearing in ACM PPoPP 2026 \cite{ppopp26-dice}} 



\category{} 
\supplement{}
\EventShortTitle{} 
\funding{} 

\RequirePackage[T1]{fontenc} 
\RequirePackage[tt=false, type1=true]{libertine} 
\RequirePackage[varqu]{zi4} 
\RequirePackage[libertine]{newtxmath}

\begin{document}
\fontsize{11}{13}\selectfont
\maketitle



\newcommand{\Acquire}{\texttt{Acquire}}         
\newcommand{\Release}{\texttt{Release}}         

\newcommand{\Nonce}{Hapax Locks} 
\newcommand{\Hapax}{Hapax Locks} 
\newcommand{\recipro}{Reciprocating Locks}


\newcommand{\LOCKEDEMPTY}{simple locked}       
\newcommand{\nullptr}{\texttt{nullptr}}         

\begin{abstract}
We present \textbf{Hapax Locks}, a novel locking algorithm that is simple, enjoys
constant-time arrival and unlock paths, provides FIFO admission order, and which is also space efficient
and generates relatively little coherence traffic under contention in the common case.  
\Hapax{} offer performance (both latency and scalability) that is comparable
with the best state of the art locks, while at the same time Hapax Locks impose
fewer constraints and dependencies on the ambient runtime environment, making them particularly 
easy to integrate or retrofit into existing systems or under existing lock
application programming interfaces
Of particular note, no pointers shift or escape ownership between threads in our algorithm.  
\footnote{source code:\url{/https://github.com/davedice/Hapax-Locks}}. 

\end{abstract}







\section{Introduction} 


\Invisible{Context; Motivation; background; frame the problem; }  
\Invisible{endow; imbue} 
 
Locks often have a crucial impact on the performance of parallel software, 
hence they remain the focus of intensive research with a 
steady stream of algorithms proposed over the last several decades. 
Picking an algorithm involves striking a balance between a number of factors,
such as : latency in the uncontended case; throughput in the contended case; admission fairness;
space requirements; coherence traffic induced by the lock; any dependencies the lock might have 
on other subsystems (such as dynamic memory allocation); and how well an algorithm might 
be retrofit under existing locking application interfaces.  
Our investigation focuses on algorithms and implementations thereof that
gracefully tolerate large numbers of dynamically created and destroyed threads and locks,
and that can be implemented easily under C++ locking interfaces and POSIX \texttt{pthreads} APIs.  
Applications expect to be able to acquire and hold multiple locks and to release
in any order -- scoped or balanced locking is not required and we have explicit \texttt{lock} and 
\texttt{unlock} operators.  
In addition, we prefer to avoid dynamic memory allocation, and the requirement
for explicit constructor and destructor methods for threads and lock instances.  
In this paper, we present \Hapax{}, which possess all aforementioned properties.  

\Invisible{possess; satisfies; provides}  
\Invisible{easy-of-use; uptake acceptability under existing APIs}  

\Invisible{Oracle Patent Disclosure Accession Number IDF-139903}  
\Invisible{Extreme uniqueness; radical uniqueness} 
\Invisible{Ephemeral evanescent ticket values; lightning; fulgar; fulmen; elding; leiptr; virtual particle } 
\Invisible{short-lived; cursory; transitory; meteoric; fugacious; momentary; emphemera; ephemeron; profluent; } 
\Invisible{unicity; unicus; uniquity}


\section{Tidex Lock Algorithm} 
Our design starts with and is inspired by the \textbf{Tidex}\cite{ppopp17-Ramalhete} mutual exclusion algorithm.  
Briefly, in Tidex, each lock instance has \texttt{Arrive} and \texttt{Depart} fields which contain thread identifier values.  
The lock is in \emph{unlocked} state if and only if \texttt{Arrive} equals \texttt{Depart}.  On arrival in the \texttt{lock} method, a
thread atomically exchanges its own unique identity value into \texttt{Arrive}, which returns the predecessor value.   
The thread then waits for \texttt{Depart} to become equal to that predecessor value that it has in hand, 
at which point it has become the owner and can enter the critical section.  In the corresponding \texttt{unlock} operator,  
the thread just stores its own thread identity value into \texttt{Depart}.  That is, a thread in \texttt{unlock}
announces that it has surrendered ownership by posting its own identity value into \texttt{Depart} -- 
that outgoing thread never knows the identity of its successor, if any exists.  
Intuitively, when a thread is waiting and its predecessor departs and deposits its identity into \texttt{Depart},
that waiting thread becomes the next owner.  
Similar to CLH\cite{craig-clh}, the queue of waiting threads is implicit, with no 
manifest or explicit in-memory linked list. 
Waiting threads, if any, know only the identity of their immediate predecessor and wait for that 
value to appear in the \texttt{Depart} field.  Collectively, however, a queue is formed.
Identifier values are expected to be temporally unique and a given value can be associated with 
at most one thread at any given time.  

\Invisible{inadvertently} 

A minor complication exists in that a thread $T1$ could inadvertently try to acquire some lock $L$, where $T1$ had recently 
acquired and released $L$, and $T1$'s identity value was still installed in $L$’s \texttt{Depart} field.  
If $T1$ were to simply atomically swap its identity value, $T1$, into $L$'s \texttt{Arrive}, then the lock 
would appear to be in \emph{unlocked} state and we might suffer exclusion failure.  
The remedy, as described in \cite{ppopp17-Ramalhete}, is that each thread should maintain 
two identity values, a primary and an alternative.  
On arrival in \texttt{lock}, a thread first fetches \texttt{Depart}, and it if finds its own primary 
identity value there, left as a residual from prior operations,  it shifts to its 
alternative ID for that \texttt{lock-unlock} episode. 

\Invisible{Remedy; Rectify; address; mitigate; fix; solution; mitigation} 
\Invisible{Narrative and derivation; origin story; explication; illustration; construction; story; creation myth; 
exposition; disquisition; evolutionary steps;  } 
\Invisible{Entailing - Contradictory - Neutral} 
\Invisible{LPU = Least publishable unit} 

\vspace{.5cm} 
\newcolumntype{?}{!{\vrule width 1pt}}
\begin{table} [h]
\centering
\begin{tabular}{ll|llll}
\toprule
Time & Action & Owner & \textbf{Arrive} & \textbf{Depart} & Waiters \\


\midrule
1   & Initial unlocked state  & - & 0 & 0 &            \\
2   & A arrives (enters)      & A & A & 0 &            \\
3   & B arrives (waits)       & A & B & 0 & B@A        \\
4   & C arrives (waits)       & A & C & 0 & C@B B@A    \\
5   & A unlocks (pass to B)   & B & C & A & C@B        \\
6   & B unlocks (pass to C)   & C & C & B &            \\
7   & C unlocks (neutral)     & - & C & C &            \\\hline
8   & C locks   (duplicate)   & C & Ĉ & C &            \\
9   & C unlocks (duplicate)   & - & Ĉ & Ĉ &            \\

\midrule[\heavyrulewidth]
\bottomrule
\end{tabular}%
\vspace{.5cm} 
\caption{Example Tidex operation - Scenario}\label{Table:Tidex}
\end{table}
\vspace{.5cm} 

Table-\ref{Table:Tidex} depicts a scenario where threads acquire and release a Tidex lock.
Initially, at time $1$, the lock is in the \emph{unlocked} state as the \texttt{Arrive} and \texttt{Depart}
values are equal.  For explication, we annotate the table with the identity of the current owner, but note that no
such field is manifest in the lock data structure.  At time $2$, thread $A$ arrives and executes 
an atomic exchange operation to install $A$ into the \texttt{Arrive} field.  The exchange returns $0$.
$A$ then fetches the \texttt{Depart} field and observes $0$, and recognizes that it has acquired
the lock without any need to wait.  $A$ enters the critical section. 
At time $3$, $B$ arrives to acquire the lock, and $B$'s exchange installs $B$ into \texttt{Depart} and
returns $A$.  $B@A$ indicates that $B$ waits for $A$ to announce its eventual departure (unlock).
At step $4$ thread $C$ arrives to acquire the lock.  $C$'s exchange returns $B$ and $C$ waits for
$B$ to announces its departure.  In step $5$, $A$ releases the lock by installing its own identity, $A$,
into the \texttt{Depart} field. As $B$ is waiting for $A$ to appear in \texttt{Depart}, $A$ has 
passed ownership to $B$, and $B$ enters the critical section.  In step $6$, $B$ releases the lock,
by storing $B$ into the \texttt{Depart} field, which is recognized by $C$.  $C$ is the new owner
and enters the critical section.  In step $7$, $C$ releases the lock by storing $C$ into
\texttt{Depart}.  At this juncture both \texttt{Arrive} and \texttt{Depart} equal $C$, so the
lock reverts to \emph{unlocked} state. 

\Invisible{Annotate; decorate; augment; } 

We continue the scenario to illustrate the requirement for alternative thread identities.
At step 8, $C$ returns to re-acquire the lock.  $C$, however, also appears as a residual value
in both \texttt{Depart} and \texttt{Arrive}.   If $C$ were to simply exchange $C$ into \texttt{Arrive},
then the lock's state would remain unchanged and we could suffer subsequent exclusion and safety failure. 
To avoid this scenario, arriving threads, before they execute the exchange into \texttt{Arrive},
must first fetch and inspect \texttt{Depart}, and, if their identity appears there, switch to 
an alternative identity, which, in this case, is $\hat C$.

\Invisible{Nothing bad -- exclusion failure -- ever happens.  Something good -- progress -- eventually happens}
\Invisible{Hapax; Neologism; Neolog; Protologism; Coin; Coinage; } 
\Invisible{Residual, Vestigial, residue} 
\Invisible{Conjure; concoct; reify; manifest; construct; fabricate; allocate;} 
\Invisible{Nonce; Token; Ticket; Unique-identifier; serial-number; inimitable number; UUID; } 
\Invisible{Post; store; deposit; install; inscribe; announce; reveal; expose; }





\lstloadlanguages{C++} 
\lstset{language=C++}
\lstset{frame=lines}
\lstset{basicstyle=\tiny\ttfamily} 
\lstset{morekeywords={nullptr}}   
\lstset{commentstyle=\itshape\color{gray}} 
\lstset{commentstyle=\slshape\color{gray}} 
\lstset{commentstyle=\itshape\color{gray}} 
\lstset{keywordstyle=\color{forestgreen}\bfseries} 
\lstset{backgroundcolor=\color{Gray95}}

\lstset{basicstyle=\fontsize{5.8}{6.5}\selectfont\ttfamily}
\lstset{basicstyle=\fontsize{7}{8}\selectfont\ttfamily}
\lstset{basicstyle=\fontsize{5}{6}\selectfont\ttfamily}
\lstset{basicstyle=\fontsize{6}{7}\selectfont\ttfamily}
\lstset{basicstyle=\fontsize{6.5}{7.5}\selectfont\ttfamily}
\lstset{basicstyle=\fontsize{7}{7.75}\selectfont\ttfamily}
\lstset{basicstyle=\fontsize{6}{7}\selectfont\ttfamily}
\lstset{basicstyle=\fontsize{5.4}{6}\selectfont\ttfamily}
\lstset{basicstyle=\fontsize{6}{6.5}\selectfont\ttfamily}       
\lstset{basicstyle=\fontsize{6.5}{7}\selectfont\ttfamily}       
\lstset{basicstyle=\fontsize{7}{7.5}\selectfont\ttfamily}       
\lstset{basicstyle=\fontsize{6}{6.5}\selectfont\ttfamily}       
\lstset{basicstyle=\fontsize{7}{7.5}\selectfont\ttfamily}       

\lstset{commentstyle=\color{blue}\itshape} 
\lstset{commentstyle=\itshape\rmfamily\color{blue}} 
\lstset{commentstyle=\itshape\rmfamily\color{gray}} 
\lstset{commentstyle=\itshape\sffamily\color{gray}} 
\lstset{commentstyle=\slshape\ttfamily\color{gray}} 
\lstset{commentstyle=\itshape\ttfamily\color{blue}} 
\lstset{commentstyle=\itshape\ttfamily\color{gray}}

\newcommand{\IRule}{\smash{\rule[-.2\baselineskip]{.4pt}{\baselineskip}\kern.5em}}

\newcommand{\TaggedN}{\mbox{}\null{}$\blacktriangleleft$} 
\newcommand{\TaggedS}{\mbox{}\hfill{}$\blacktriangleleft$} 
\newcommand{\TaggedTT}{\mbox{}\hfill{}$\blacktriangleleft\blacktriangleleft\blacktriangleleft$} 
\newcommand{\TaggedRHS}{\hfill\hspace{-1\textwidth}$\blacktriangleleft$} 
\newcommand{\TaggedCR}{\vspace*{-\baselineskip}\\$\blacktriangleleft$} 
\newcommand{\TaggedVSAfter}{\vspace*{-.5\baselineskip}\\$\blacktriangleleft$} 
\newcommand{\TaggedVSBefore}{\\\vspace*{-.5\baselineskip}$\blacktriangleleft$} 
\newcommand{\TaggedLL}{\llap{$\blacktriangleleft$}} 
\newcommand{\TaggedTN}{\\\vspace*{-.7\baselineskip}$\blacktriangleleft$\\} 
\newcommand{\TaggedTR}{\vspace{-.5\baselineskip}\\\raisebox{+.1\baselineskip}{$\blacktriangleleft$}} 
\newcommand{\TaggedSmash}{\vspace{-.6\baselineskip}\\\smash{\strut$\blacktriangleleft$}} 
\newcommand{\TaggedHSpace}{\hspace{-1\linewidth}$\blacktriangleleft$} 
\newcommand{\TaggedTViableA}{\vspace{-.5\baselineskip}\\$\blacktriangleleft$} 
\newcommand{\TaggedTViableB}{\vspace{-.6\baselineskip}\\$\blacktriangleleft$\\} 
\newcommand{\TaggedTViableC}{\vspace{-.6\baselineskip}\hfill\break$\blacktriangleleft$} 
\newcommand{\TaggedTViableD}{\vspace{-.6\baselineskip}\newline$\blacktriangleleft$} 
\newcommand{\TaggedTViableE}{\vspace{-.6\baselineskip}\newline$\blacktriangleleft$} 
\newcommand{\TaggedTViableF}{\vspace{-.5\baselineskip}\newline\makebox[0pt][l]{$\blacktriangleleft$}} 
\newcommand{\TaggedTViableG}{\vspace{-.6\baselineskip}\newline$\blacktriangleleft$} 
\newcommand{\TaggedT}{}

\newcommand{\HIHI}{\makebox[0pt][l]{\color{cyan}\rule[-4pt]{0.65\linewidth}{14pt}}}
\newcommand{\HICCFailA}{\makebox[0pt][l]{\color{cyan!25}\rule[-.3\baselineskip]{\widthof{\hfill}}{7pt}}}
\newcommand{\HICCViable}{\makebox[0pt][l]{\color{cyan!25}\rule[-.3\baselineskip]{0.65\textwidth}{7pt}}}
\newcommand{\HICC}{\makebox[0pt][l]{\color{cyan!25}\rule[-.3\baselineskip]{0.65\textwidth}{7pt}}}
\newcommand{\RaisedRule}[2][0em]{\leaders\hbox{\rule[#1]{1pt}{#2}}\hfill}

\newcommand{\Global}[1]{\fboxsep0.5pt\colorbox{green!40}{#1}}  
\newcommand{\FastPath}[1]{\fboxsep0.5pt\colorbox{purple!40}{#1}}  
\newcommand{\HandOver}[1]{\fboxsep0.5pt\colorbox{blue!60}{#1}}         

\Invisible{Annotate; Decorate; Mark; Tag} 

Tidex is related to CLH\cite{craig-clh} and HemLock\cite{spaa21-Dice} in that contending threads wait 
on state related to the \emph{predecessor} value 
that was returned from the atomic swap executed on arrival.
The arrival ``doorway'' phase \cite{cacm74-lamport} and unlock paths run in constant-time.  
While simple, the key downsides are the need on arrival to fetch and check the \texttt{Depart} field, 
and, most critical for scalability, the use of global spinning by waiters on the \texttt{Depart} field.

In Ticket locks\cite{tocs91-MellorCrummey}, in contrast, a thread waits for its \emph{own} ticket value to
appear in the lock's \texttt{Grant} (``now serving'') field, whereas in Tidex  
a thread waits for its predecessor's identity, obtained via the atomic exchange,
to appear in a variable associated with the lock.

For reference, we provide an example Tidex Lock implementation in modern C++ in Listing-\ref{Listing:Tidex}.
To assist the reader, we have annotated accesses to shared global locations in \Global{green}.  
For brevity, this version expresses the critical section as a C++ lambda expression.
Non-escaping lambdas are efficient and add no particular runtime overhead
\footnote{Example usage : \texttt{Tidex L \{\};\allowbreak{} int v = 5; \allowbreak{}L + [\&]\{ v += 2;\} ;}}. 
The \texttt{assert} statements in the listing document \Exclude{and express} invariants and do not constitute runtime checks against 
errant application usage of the lock.
Critically, they are not required for correct operation and will be disabled in normal non-debug builds. 
To achieve the desired progress properties, we expect that C++ 64-bit \texttt{std::atomic} load, store, \texttt{fetch\_add} 
and \texttt{compare\_and\_swap} run in constant-time, and, specifically, are not emulated via \texttt{load-locked} and \texttt{store-conditional} loops.  
We also assume the existence of a \texttt{Pause()} operator for polite busy waiting.

\Invisible {concessions to brevity} 
\Invisible {Explicate, explain, expound, illustrate, demonstrate, show, illuminate, depict} 
\Invisible {Narrative, annotated scenario, flow, example}

\lstset{caption={Tidex Lock Algorithm}}
\lstset{label={Listing:Tidex}}
\begin{lstlisting}[mathescape=true,escapechar=\%]
// Variant of Tidex Lock algorithm -- related to CLH and Ticket Locks
// https://doi.org/10.1145/2851141.2851171
// PPoPP 2016 Pedro Ramalhete and Andreia Correia

struct Tidex {
 %\IRule% std::atomic <uintptr_t> Arrive {0} ;       // ingress -- most recent arrival
 %\IRule% std::atomic <uintptr_t> Depart {0} ;       // egress  -- most recent departure

 %\IRule% // We desire "lock_free" for progress properties and performance, 
 %\IRule% // not correctness
 %\IRule% static_assert (std::atomic<uintptr_t>::is_always_lock_free) ; 
 %\IRule% 
 public : 
 %\IRule% inline auto %{\fboxsep1pt\colorbox{yellow!50}{operator+}}%(std::invocable auto && csfn) %$\rightarrow$% void {
 %\IRule%  %\IRule% // Acquire() : 
 %\IRule%  %\IRule% // Conjure unique thread identity value 
 %\IRule%  %\IRule% // We leverage the address of TSelf as a temporally unique identity 
 %\IRule%  %\IRule% // and never access the contents of the array.  
 %\IRule%  %\IRule% // Element [0] is the primary identity and [1] is the alternative.
 %\IRule%  %\IRule% static constinit thread_local char TSelf [2] ; 
 %\IRule%  %\IRule% auto ISelf = uintptr_t(TSelf) ; 
 %\IRule%  %\IRule%  
 %\IRule%  %\IRule% ISelf += %\Global{Depart}%.load (std::memory_order_acquire) == ISelf ; 
 %\IRule%  %\IRule% auto prv = %\Global{Arrive}%.exchange (ISelf) ; 
 %\IRule%  %\IRule% assert (prv %$\neq$% ISelf) ; 
 %\IRule%  %\IRule% while (%\Global{Depart}%.load (std::memory_order_acquire) %$\neq$% prv) {
 %\IRule%  %\IRule%  %\IRule% Pause() ; 
 %\IRule%  %\IRule% }
 %\IRule%  %\IRule%  
 %\IRule%  %\IRule% %{\rule[0pt]{0.65\linewidth}{.5pt}}%
 %\IRule%  %\IRule% EnterCS:
 %\IRule%  %\IRule% // We pass ISelf as context from lock to the corresponding 
 %\IRule%  %\IRule% // unlock operation
 %\IRule%  %\IRule% // In sustained steady-state contention, we expect just one thread
 %\IRule%  %\IRule% // to have arrived during the execution of the CS.
 %\IRule%  %\IRule% // Arrival and depart rates should eventually equilibrate
 %\IRule%  %\IRule% assert (Depart.load() %$\neq$% Arrive.load() ; 
 %\IRule%  %\IRule% csfn() ; 
 %\IRule%  %\IRule% assert (Depart.load() %$\neq$% Arrive.load() ; 
 %\IRule%  %\IRule% %{\rule[0pt]{0.65\linewidth}{.5pt}}%
 %\IRule%  %\IRule%  
 %\IRule%  %\IRule% // Release() :
 %\IRule%  %\IRule% assert (Depart.load() %$\neq$% ISelf) ; 
 %\IRule%  %\IRule% %\Global{Depart}%.store (ISelf, std::memory_order_release) ; 
 %\IRule% }
} ; 

\end{lstlisting}  

\section{Hapax Locks}


\Hapax{} are based on, but address the aforementioned performance shortcomings inherent in Tidex Locks.  
With \Hapax{}, threads no longer bother to maintain their own ID and alternative ID values.  
In \texttt{lock}, our thread \emph{notionally} executes an atomic \texttt{fetch\_add(1)} against a global shared 64-bit ID generator 
to conjure a new unique 64-bit value, and we use that as the thread ID for the acquire-release episode.    
The single generator instance is used for all threads and all locks in a process.  
The value from the generator is our \textbf{hapax}\footnote{
In linguistics, \emph{hapax legomenon}, -- or just \emph{hapax} (\textalpha\textpi\textalpha\textxi $=$ ``once'') -- 
refers to a word that appears just once in a body of text.
In Hapax Locks, a particular hapax value is a \emph{nonce} and appears at most once in the execution history of a process.
} value.   
Instead of a thread just having two Tidex-style identity values, it can now have multiple, allocated dynamically,
one unique for each lock acquisition operation.  
This approach avoids any need to fetch and check \texttt{Depart} to ensure that our value does not happen to already appear in 
\texttt{Arrive} or \texttt{Depart}.   A given hapax value, once allocated and made public by installation
into \texttt{Arrive}, is never reused or recycled in the history of the process.  
With 64-bit values, we don’t, as a practical concern, need to worry about 
overflow and wrap-around of the hapax values generated.   
Unique hapax values are cheap to generate, plentiful, and for all practical purposes, limitless.   


\Invisible{Hapax values are unique over the complete history of the process} 
\Invisible{Changed name under advisement} 

To avoid the global hapax identity value generator becoming a coherence hot spot,  in practice, we allocate \emph{blocks} (consecutive runs) 
of unique hapax identifiers to a thread.  Once it has allocated a block, our thread can then locally sub-allocate (just by incrementing the value) 
until the locally held block is exhausted, at which case it must reprovision from the global ID generator.   
The values generated from the local block are still globally unique \footnote{
We note in passing that unique hapax values have other applications.  In the TWA locking algorithm\cite{EuroPar19-TWA}, 
for instance, we can replace an atomic increment (Listing-1.1 line 71) with the store of a fresh hapax value.}. 
We argue this still avoids any practical real-world roll-over or wrap around concerns for the values.
By reducing traffic on the ID generator, we avoid a scalability impediment.  
In particular, large blocks reduce the frequency of access to the global block allocator.  
In the attached code, we allocate via 64K blocks of hapax values, so the upper 48 bits of a hapax 
are associated with a specific thread and the lower 16-bit are used for local sub-allocation by that thread.   
A thread needs to reprovision from the global allocator every 64K lock acquisition operations. 
Developers can tune the block size as they see fit.  
 
Instead of a using a single global allocator, we observe
that more sophisticated and scalable allocators are possible.  For instance, using an array of allocator ``lanes'', 
offset appropriately, and selected randomly, or by geographic location in the system topology, will
yield more scalable allocation, but such approaches were not found profitable as the allocation
is amortized over the block size.


We note that threads might come into being, allocate a block, 
and then die, which, absent any other recovery mechanism, would cause all the remaining hapax values in 
that thread's sub-block to be abandoned and discarded.  In turn, such behavior might cause the global block 
allocator to overflow prematurely.  As such, excessively large block values may not be appropriate.  
We believe a reasonable compromise to be 64K blocks with a 48+16 split.  

Finally, instead of allocating blocks of hapax values per-thread, we could instead use per-CPU allocation 
via techniques based on\emph{multi-processor restartable critical sections} (RCS)
\cite{ismm02-dice,MP-RCS-PLABS,RCS-patent,vee05-dice,blog-khuong-restartable},
still avoiding atomic operations for fast-path allocation within a block.  
RCS presages the linux \emph{RSEQ} facility\cite{RCS-1,RCS-2}.


\Invisible{Embellishment : array of randomly selected Nonce allocators, picked at random and offset accordingly.} 
\Invisible{Take inspiration from; borrow a solution from;} 

Both Tidex and Ticket locks use global spinning, which can result in potentially excessive write invalidation rates
\footnote{Imagine a contended Ticket Lock with 10 waiting threads, and no arriving threads.  
When the owner surrenders the lock, the write will invalidate, at worst case, in 10 remote caches. 
Only one of those waiting threads that suffered invalidation will become the next owner and
the remainder will continue to busy-wait, pulling the line back into their caches after the invalidation.  
When that new owner ultimately surrenders ownership, its write will invalidate in 9 remote caches, and so on.
So if we have $T$ waiting threads, each with a private multiple-reader-exclusive-writer MESI-style cache, 
then the number of invalidations required to pass ownership to all the threads, assuming no new arrivals,
is at least $(T^2)/2$ -- potentially quadratic write invalidation rates.  We refer to the number of caches that require invalidation from a given
store instruction as the \emph{invalidation set} or \emph{invalidation blast zone} of that store.
Broadly, larger \emph{invalidation sets} may incur more latency and consume more interconnect bandwidth.}.  
In Tidex, for instance, all waiting threads for a given lock busy wait on the \texttt{Depart} field.  
To avoid such global spinning, we take inspiration from TWA\cite{EuroPar19-TWA,arxiv-TWA} -- itself derived from ticket locks -- 
and implement a shared \emph{waiting array}, where
all threads and all locks in a process use the same shared array.   
We opted to provision the array with 4096 slots.
In theory, the number of slots should be a function of the number of logical processors 
in the system, but 4096 is reasonable even on very large modern systems. 
Threads perform long-term waiting via elements in the array.   
TWA hashed the ticket value and lock address to form an index into the waiting array, and the elements in those array slots 
were slot-specific sequence numbers.  When TWA waiters observe the slot sequence number change, they would then recheck 
the actual ticket lock \texttt{Grant} ``now serving'' field to see if they were granted ownership.  We refer to this as \emph{proxy waiting}. 
Updates to the array elements constitute a conservative hint and notification to waiters that the actual location of interest may 
have been modified -- the waiter is then expected to check the authoritative ``ground truth'' variable.  
In the TWA \texttt{unlock} operator, after updating \texttt{Grant}, TWA then atomically increments the corresponding slot sequence number 
to signal any waiters.  
Note, however, that because of collisions, waiters observing the sequence number change can not safely 
intuit that \texttt{Grant} had shifted to the waited-for value, so waiters are required to conservatively
recheck \texttt{Grant} and, if necessary, resume the proxy waiting protocol.
Absent hash collisions, we have \emph{private waiting}, but as collisions may manifest, 
we say we have \emph{semi-private waiting}.  

\Invisible{CREW = Concurrent-read-exclusive-write} 
\Invisible{Invalidation radius; invalidation blast radius; invalidation blast zone; invalidation victim set;} 
\Invisible{Birthday paradox relates to collision rate} 

For Hapax Locks, we could just apply any general hash function, taking the hapax value and, optionally, the lock 
address, to form an index into the waiting array.  That would certainly suffice, but a better option is available.    In the mapping function 
from hapax to waiting array index, we extract the high-order bits in the hapax, which are associated with the thread,
and discard the rest (the offset into the thread's block), 
and then apply a simple cache-aware lightweight hash to form the index.   
(We multiply by 17\cite{WeylSequence} to map adjacent block zones onto 
different cache sectors or lines in the array, to reduce false sharing).   
So, for better performance, the hash — mapping function from waited-upon hapax 
value to waiting array index — is intentionally aware of the block allocation scheme.    
While not strictly required, we believe this is a useful optimization and strikes a balance
between temporal cache locality (re-use by the same thread) in the array and the risk of collisions
or near collisions (false sharing) between threads. 
This approach is illustrated in the \texttt{ToSlot} function in Listing-\ref{Listing:Hapax-Locks}.  

In the \texttt{unlock} path, a thread writes its hapax — the one associated with that specific acquire-release 
episode -- into \texttt{Depart}, and then also into the associated slot in the waiting array.   
This outgoing thread posts notification into the waiting array anticipating that its immediate successor,
if any, will observe the change and recognize that it has been granted ownership.  
(If the successor observes the exact waited upon value appear in the slot, it can return expeditiously without needing to
check \texttt{Depart}, otherwise if observes the value in the array change to any other value, it must conservatively
recheck \texttt{Depart}).  
We require only simple C++ atomic loads and stores in the waiting array slots, unlike TWA, 
where updates needed to atomically increment the slot element.   
If desired, as an optional optimization, we can favor uncontended operation, and, in \texttt{unlock}, 
after storing into \texttt{Depart}, 
check if there are any newly arrived waiters, by fetching \texttt{Arrive}, and potentially skip the waiting slot update when
there are known to be no waiting threads. 

As noted above, if a waiter happens to see its waited-upon hapax value appear in the array, it does \emph{not} need to recheck the 
\texttt{Depart} value to confirm handover, but, rather it can directly enter the critical section.  
We call this mode of transfer \emph{direct expedited handover}.  
This technique is sound because the hapax values are globally unique and non-recurring.
Crucially, this optimization could \emph{not} be safely applied in TWA.  
A hapax value conceptually maps uniquely to a thread, a lock and a specific acquire episode by that thread on that lock.  
So in the normal case of no collisions in the array, we can effect a very fast direct handover where the recipient 
-- the immediate successor -- 
doesn’t need to check and confirm the \texttt{Depart} value.  This, in turn, reduces coherence traffic, and makes handover recognition
more efficient and reduces path latency. 
In the case of collisions in the array, a waiter might see the slot value change to a non waited-upon value.  
In that case it needs to recheck the actual \texttt{Depart} field value, 
to determine whether it is or is not the new owner.   The waiter's hapax value-of-interest could have flickered into visibility and 
then have been ``stomped'' or overwritten by unrelated concurrent actions, but not to the waited-upon value.  
So when a waiter observes the slot change 
to a value not of interest, it needs to recheck the definitive or authoritative ``ground truth'', which is value
in the lock's \texttt{Depart} field.  That scenario should be rare and only manifest via collisions in the array.     
Crucially, assuming hapax values never overflow and wrap around (which would compromise correctness)
no \emph{ABA}\cite{ArtOf} pathologies are possibly given the non-recurring nature of hapax values --
a slot will never flicker back to an old value, which, could potentially leave stranded waiters.  

\Invisible{stomp; overwritten; flicker; emphemeral; transient} 

Under contention, the value of the current outgoing owner's \texttt{hapax} variable,
executing in \texttt{unlock}, is the same as the incoming immediate successor's \texttt{pred} variable, as found
in the successor's \texttt{lock} phase.
In Ticket locks, a thread knows that its immediate waiting successor must hold a ticket value
of just one more than the ticket value allocated to the current owner.  In Tidex and Hapax Locks, however,
the departing thread has no particular knowledge about the successor, other than being able to determine if it exists.  




\Invisible{Throw-away; one-shot; one-time; disposable; single-use; non-retrograde}

\lstset{caption={Hapax Locks}}
\lstset{label={Listing:Hapax-Locks}}
\begin{lstlisting}[mathescape=true,escapechar=\%]
struct HapaxLock { 
 %\IRule% struct Slot { std::atomic<uint64_t> Notification {0} ; } ; 
 %\IRule%  
 %\IRule% // Map a hapax identity value to a slot in the global waiting array
 %\IRule% // We multiply by 17 as 17 is coprime with ArraySize, allowing us to fully 
 %\IRule% // utilize all slots in the array assuming a dense set of zone values.
 %\IRule% // Using 17 also causes adjacent thread zone values to fall onto slots in 
 %\IRule% // different cache sectors, reducing false sharing
 %\IRule% // See also :
 %\IRule% // https://en.wikipedia.org/wiki/Low-discrepancy_sequence
 %\IRule% // https://en.wikipedia.org/wiki/Weyl_sequence
 %\IRule% // https://en.wikipedia.org/wiki/Equidistributed_sequence
 %\IRule% // We add the optional "salt" value to help reduce the odds of collisions
 %\IRule% // between locks that might move in unison and to lessen the
 %\IRule% // impact of "multi-waiting" as found in HemLock. 
 %\IRule% // A well-chosen hash reduces both true collisions and near/proximal
 %\IRule% // collisions that might result in false sharing.  
 %\IRule%  
 %\IRule% Slot * ToSlot (uint64_t hapax) { 
 %\IRule%  %\IRule% static constexpr int ArraySize = 4096 ; 
 %\IRule%  %\IRule% static_assert (ArraySize > 0 && (ArraySize & (ArraySize-1)) == 0) ; 
 %\IRule%  %\IRule% static Slot WaitingArray alignas(4096) [ArraySize] {0} ; 
 %\IRule%  %\IRule% auto salt = uint32_t(uintptr_t(this)) ; 
 %\IRule%  %\IRule% uint32_t ix = ((salt + uint32_t(hapax >> 16)) * 17) & (ArraySize-1) ;
 %\IRule%  %\IRule% return WaitingArray + ix ; 
 %\IRule% }
 %\IRule%  
 %\IRule% // Arrive is the hapax identity of the most recently arrived thread.
 %\IRule% // Depart is the hapax identity of the most recently departed thread.
 %\IRule% // That is, the last thread to release the lock.
 %\IRule% // The lock is held iff Arrive == Depart.  
 %\IRule% // The owner is implicit. 
 %\IRule% // Placement choice : sequester or collocate Arrive and Depart fields
 %\IRule% std::atomic <uint64_t> Arrive {0} ;       // ingress
 %\IRule% std::atomic <uint64_t> Depart {0} ;       // egress
 %\IRule%  
 %\IRule% static inline constinit thread_local uint64_t PrivateHapax = 0 ; 
 %\IRule% static inline constinit std::atomic<uint64_t> HapaxAllocator alignas(128) {0} ; 
 %\IRule%  
 public : 
 %\IRule% inline auto %{\fboxsep1pt\colorbox{yellow!50}{operator+}}%(std::invocable auto && csfn) %$\rightarrow$% void {
 %\IRule%  %\IRule% // Acquire the lock ...
 %\IRule%  %\IRule% // First, conjure a unique hapax identity for this specific 
 %\IRule%  %\IRule% // acquire-release episode. 
 %\IRule%  %\IRule% // The hapax is globally and temporally unique and specific to 
 %\IRule%  %\IRule% // this thread, this lock, and this episode
 %\IRule%  %\IRule% auto hapax = PrivateHapax ++ ; 
 %\IRule%  %\IRule% if ((hapax & 0xFFFF) == 0) [[unlikely]] { 
 %\IRule%  %\IRule%  %\IRule% // crossed edge of block allocation 
 %\IRule%  %\IRule%  %\IRule% // current block is exhausted so must reprovision
 %\IRule%  %\IRule%  %\IRule% // In this particular example the high 48-bits of the 64-bit hapax 
 %\IRule%  %\IRule%  %\IRule% // encode the thread "zone" and the lower 16 are the sub-sequence 
 %\IRule%  %\IRule%  %\IRule% // from which the thread can allocate locally.  
 %\IRule%  %\IRule%  %\IRule% hapax = HapaxAllocator.fetch_add(1)+1 ; 
 %\IRule%  %\IRule%  %\IRule% assert (hapax %$\neq$% 0) ; 
 %\IRule%  %\IRule%  %\IRule% hapax = hapax << 16 ; 
 %\IRule%  %\IRule%  %\IRule% assert (hapax > PrivateHapax) ; 
 %\IRule%  %\IRule%  %\IRule% PrivateHapax = hapax + 1 ; 
 %\IRule%  %\IRule% }
 %\IRule%  %\IRule% assert (hapax %$\neq$% 0) ;       // by convention, 0 is reserved   
 %\IRule%  %\IRule% assert (hapax %$\neq$% Depart.load ()) ; 
 %\IRule%  %\IRule% assert (hapax %$\neq$% Arrive.load ()) ; 
 %\IRule%  %\IRule% auto pred = %\Global{Arrive}%.exchange (hapax) ; 
 %\IRule%  %\IRule% assert (pred %$\neq$% hapax) ; 
 %\IRule%  %\IRule%  
 %\IRule%  %\IRule% // The value in Depart is the authoritative and definitive ground truth 
 %\IRule%  %\IRule% // as to the lock state and whether hand-over has been effected.  
 %\IRule%  %\IRule% // At any given time, at most one waiting thread, the immediate 
 %\IRule%  %\IRule% // successor, will be waiting for a given hapax to appear in the 
 %\IRule%  %\IRule% // Depart field,  at which time that successor knows ownership has 
 %\IRule%  %\IRule% // been transfered.  
 %\IRule%  %\IRule% // To safely busy wait on the proxy slots and close any races vs 
 %\IRule%  %\IRule% // the concurrent predecessor thread running in the unlock() phase, 
 %\IRule%  %\IRule% // we use the following protocol :
 %\IRule%  %\IRule% // (1)  Fetch ToSlot(pred)%$\rightarrow$%Notification into the local LastSeen variable
 %\IRule%  %\IRule% // (2)  Fetch Depart to ensure we have not yet been consigned 
 %\IRule%  %\IRule% //      ownership and should continue to wait.  
 %\IRule%  %\IRule% //      That is, ratify and confirm via Depart that we should still 
 %\IRule%  %\IRule% //      wait and have not yet been granted ownership.  
 %\IRule%  %\IRule% // (3)  Busy-wait on ToSlot(pred)%$\rightarrow$%Notification while it remains equal 
 %\IRule%  %\IRule% //      to LastSeen 
 %\IRule%  %\IRule% // Complementary code in the unlock() phase updates Depart first and 
 %\IRule%  %\IRule% // then the slot.  
 %\IRule%  %\IRule% //
 %\IRule%  %\IRule% // It is highly likely that Depart remains resident in our local cache
 %\IRule%  %\IRule% // between the 1st and 2nd iteration.
 %\IRule%  %\IRule% // Under contention, we expect that the 1st iteration will incur a
 %\IRule%  %\IRule% // coherence miss on Depart,  but the remainder of the locking episode 
 %\IRule%  %\IRule% // will not usually miss on Depart. 
 %\IRule%  %\IRule% 
 %\IRule%  %\IRule% uint64_t LastSeen = 0 ;         // most recently observed slot value
 %\IRule%  %\IRule% while (%\Global{Depart}%.load (std::memory_order_acquire) %$\neq$% pred) { 
 %\IRule%  %\IRule%  %\IRule% // Contention : slow-path waiting 
 %\IRule%  %\IRule%  %\IRule% // Primary busy-wait waiting loop ... 
 %\IRule%  %\IRule%  %\IRule% // Pred is only 0 on the initial locking request
 %\IRule%  %\IRule%  %\IRule% // and we can't have contention on the 1st request, thus pred %$\neq$% 0 
 %\IRule%  %\IRule%  %\IRule% assert (pred %$\neq$% 0) ; 
 %\IRule%  %\IRule%  %\IRule% auto Verify = LastSeen ; 
 %\IRule%  %\IRule%  %\IRule% ProxyWaiting : 
 %\IRule%  %\IRule%  %\IRule% for (;;) { 
 %\IRule%  %\IRule%  %\IRule%  %\IRule% LastSeen = ToSlot(pred)%$\rightarrow$%%\Global{Notification}%.load (std::memory_order_acquire) ; 
 %\IRule%  %\IRule%  %\IRule%  %\IRule% if (LastSeen == pred) { 
 %\IRule%  %\IRule%  %\IRule%  %\IRule%  %\IRule% // Try to detect expedited direct handover which lets us skip 
 %\IRule%  %\IRule%  %\IRule%  %\IRule%  %\IRule% // coherence traffic otherwise incurred by re-fetching Depart
 %\IRule%  %\IRule%  %\IRule%  %\IRule%  %\IRule% assert (Depart.load() == pred) ; 
 %\IRule%  %\IRule%  %\IRule%  %\IRule%  %\IRule% goto EnterCS ; 
 %\IRule%  %\IRule%  %\IRule%  %\IRule% }
 %\IRule%  %\IRule%  %\IRule%  %\IRule% if (LastSeen %$\neq$% Verify) [[unlikely]] break ;  // recheck Depart 
 %\IRule%  %\IRule%  %\IRule%  %\IRule% Pause() ; 
 %\IRule%  %\IRule%  %\IRule% }
 %\IRule%  %\IRule% } 
 %\IRule%  %\IRule%  
 %\IRule%  %\IRule% %{\rule[0pt]{0.65\linewidth}{.5pt}}% 
 %\IRule%  %\IRule% EnterCS: 
 %\IRule%  %\IRule% assert (Depart.load() %$\neq$% Arrive.load()) ;
 %\IRule%  %\IRule% assert (Depart.load() == pred) ; 
 %\IRule%  %\IRule% // Execute the critical section, expressed as a lambda  
 %\IRule%  %\IRule% // Pass hapax value as context from lock to the corresponding unlock
 %\IRule%  %\IRule% // In sustained steady-state contention, we expect just one new thread
 %\IRule%  %\IRule% // to arrive during the CS 
 %\IRule%  %\IRule% // Arrival and depart rates should equilibrate
 %\IRule%  %\IRule% csfn() ;
 %\IRule%  %\IRule% assert (Depart.load() %$\neq$% Arrive.load()) ; 
 %\IRule%  %\IRule% assert (Depart.load() == pred) ; 
 %\IRule%  %\IRule% %{\rule[0pt]{0.65\linewidth}{.5pt}}% 
 %\IRule%  %\IRule%  
 %\IRule%  %\IRule% // Unlock() ...
 %\IRule%  %\IRule% assert (Depart.load() %$\neq$% hapax) ; 
 %\IRule%  %\IRule% %\Global{Depart}%.store (hapax, std::memory_order_release) ; 
 %\IRule%  %\IRule%  
 %\IRule%  %\IRule% // Poke the long-term waiting slot associated with hapax value
 %\IRule%  %\IRule% // Post the notification into waiting array
 %\IRule%  %\IRule% // This informs waiters that we are renouncing or relinquishing 
 %\IRule%  %\IRule% // ownership
 %\IRule%  %\IRule% assert (ToSlot(hapax)%$\rightarrow$%Notification.load() %$\neq$% hapax) ; 
 %\IRule%  %\IRule% ToSlot(hapax)%$\rightarrow$%%\Global{Notification}%.store (hapax, std::memory_order_release) ; 
 %\IRule% } 
} ;


\end{lstlisting}  

In Listing-\ref{Listing:Hapax-Locks} we show an implementation of the Hapax Locks in modern C++.


\section{Hapax Lock Variation with Visible Waiters} 

In \Hapax{} presented in Listing-\ref{Listing:Hapax-Locks}, above, when releasing a lock, the lock holder
must store its hapax into both the \texttt{Depart} field, and then into the corresponding waiting array slot.  
(More precisely, the store into the slot must not
become visible before the store into \texttt{Depart}). 
Each of these stores can incur a cache miss and prolong the lock handover transition duration, in turn
limiting scalability\cite{isca10-eyerman}.  In this section,
we describe an optimization that eliminates one of those stores in the common case, yielding an even
more scalable variant.  It achieves this by using \emph{visible waiters} which in turn enables 
\emph{assured positive handover}, which allows the thread executing in \texttt{unlock} to skip
the store into \texttt{Depart} if it can confirm passage of ownership to its successor. 

We say the baseline version of \Hapax{} shown in Listing-\ref{Listing:Hapax-Locks} uses \emph{invisible waiters}
as waiters only monitor the waiting array, but never write into the array, whereas in the version
described in this section, with \emph{visible waiters}, waiters explicitly announce their existence 
by updating slots in the waiting array.  

In Listing-\ref{Listing:Nonce-Visible-Waiters}, 
accesses to shared globals in the contented fast path are annotated in \colorbox{purple!40}{purple}
and the key fast-path assured positive handover accesses are shown in \colorbox{blue!40}{blue}.  
These reflect the dominant mode of execution under sustained contention.
Critically, the atomic CAS (compare-and-swap) operation in \texttt{unlock} efficiently and expeditiously transfers ownership 
to its waiting successor running in \texttt{lock}.  



\Invisible{Dominant hot path}  


By convention, we ensure a hapax value of 0 is never allocated.  
Normally slots in the array are 0 indicating they are unoccupied.  Threads arriving to wait first try to CAS 
the slot associated with the waited-upon value (which is non-0) from 0 to that value.  
If that CAS fails, because the slot is occupied by another waiter, they then revert to degenerate 
Tidex-style global spinning.  This should be rare and only occur on hash collisions in the array.   
Otherwise, that waiting thread owns and has reserved the slot for the duration of its waiting phase and then busy waits 
for the value to change to \emph{any} other value.  The  corresponding \texttt{unlock} operator 
tries to clear the slot back to 0 with CAS.  The slot value can then \emph{flicker} to 0 and then 
to other values, because of concurrent activities, so the reader (waiter) is safe to interpret 
\emph{any} change in the value as indicating that ownership has been conveyed.  
This protocol is safe by virtue of the hapax value non-recurring property.   
The slot can not ever revert to a previous value, so we do not miss wakeups to a waiter.  

\Invisible{This form is our preferred embodiment, and provide the best throughput
under contention.} 
\Invisible{Preferred; ultimate; best-of; candidate; } 
\Invisible{Assured; positive; certainty; certain; affirmative; definite; express; conclusive; affirm; surety;  
inexorable; ineluctable; guaranteed; warranteed; ensured; concurrence; accord; handshake; concordance; mutuality; } 

This variation effects contended lock ownership handover via 2 distinct modes : 
\begin{enumerate*}[label={\textbf{(\alph*)}}]
\item The outgoing thread, executing in the unlock phase, uses an atomic compare-and-swap (CAS),
on the associated slot in the array to attempt to overwrite its own hapax value with 0.  
The particular slot in the array is determined via a hash on the hapax value.  
Recall that the value of the current outgoing owner's \texttt{hapax} variable, in \texttt{unlock}, is 
the same as the incoming successor's \texttt{pred} variable in the successor's lock phase.  
If the CAS is successful, the outgoing thread has accomplished expedited handover 
and foregoes updating the \texttt{Depart} field.  
The handoff is \emph{assured} (\emph{positive}) as, by virtue of the CAS, 
the departing thread knows with certainty that the waiter exists, and is
waiting at the slot, and will be able to observe the change, providing a synchronous
rendezvous between the departing thread and that successor. 
The waiting successor detects positive handover by observing that the hapax value it 
installed in the array, shifts, by virtue of the CAS by the predecessor, and can avoid
checking the \texttt{Depart} field, and can directly enter the critical section.  
Such expedited positive handover -- accelerated departure -- is the preferred mode of succession, 
enjoying the shortest paths and least latency. 
Critically, under sustained contention, neither the outgoing thread, running in \texttt{unlock}, nor the incoming successor,
waiting in \texttt{lock}, access any locations in the shared lock body; \Hapax{} shares this desirable attribute
with MCS\cite{tocs91-MellorCrummey} and CLH\cite{craig-clh,ipss94-magnusson}.  
The outgoing thread, in \texttt{unlock}, does not need to store into \texttt{Depart} and the
incoming successor, in \texttt{lock}, also does not need to fetch \texttt{Depart} to confirm handover, so
we reduce coherence traffic on cache lines underlying the lock body.  
\texttt{Unlock} only needs to update the \texttt{Depart} field when there is no contention or
if there were collisions in the waiting array.  
\item The attempt at positive expedited handover fails -- the CAS fails -- so the outgoing thread
reverts to the Tidex protocol and updates the \texttt{Depart} field, which is observed by the successor.   
Expedited positive handover can fail because of hash collisions on hapax values in the array, 
or if the unlock operation commences before the tardy successor has had a chance to register 
and make itself visible in the waiting array.  The \texttt{unlock} operation can not
determine with certainty if a successor exists -- if the lock is contended or not -- so 
it reverts to a more conservative protocol to release the lock. 
\end{enumerate*} 

\lstset{caption={Hapax Lock Variation with Visible Waiters}}
\lstset{label={Listing:Nonce-Visible-Waiters}}
\begin{lstlisting}[mathescape=true,escapechar=\%]
// HapaxLock with visible waiters
//
// @  Visible waiters in waiting array slots
// @  Each visible waiter occupies a slot in the waiting array
// @  Amenable to futex address-based waiting
// @  Falls back to either classic Tidex or HapaxLock in the event of hash 
//    collisions in the waiting array
// @  Normally slots in the array are 0

struct HapaxVW { 
 %\IRule% using Hapax = uint64_t ; 
 %\IRule%  
 %\IRule% // Hapax value of singleton registered "visible" waiter 
 %\IRule% struct Slot { Atomic <uint64_t> VisibleWaiter {0} ; } ; 
 %\IRule%  
 %\IRule% Slot * ToSlot (uint64_t hapax) { 
 %\IRule%  %\IRule% static constexpr int ArraySize = 4096 ; 
 %\IRule%  %\IRule% static Slot WaitingArray alignas(4096) [ArraySize] {0} ; 
 %\IRule%  %\IRule% // Add optional "salt" into hash function to reduce undesirable 
 %\IRule%  %\IRule% // collisions between unrelated contended locks that might move in 
 %\IRule%  %\IRule% // unison and to less the impact of "multi-waiting" as found in HemLock. 
 %\IRule%  %\IRule% auto salt = uint32_t(uintptr_t(this)) ; 
 %\IRule%  %\IRule% uint32_t ix = ((salt + uint32_t(hapax >> 16)) * 17) & (ArraySize-1) ;
 %\IRule%  %\IRule% return WaitingArray + ix ; 
 %\IRule% }
 %\IRule%  
 %\IRule% std::atomic <uint64_t> Arrive {0} ;   // ingress = most recently arrived
 %\IRule% std::atomic <uint64_t> Depart {0} ;   // egress = most recently departed
 %\IRule%  
 %\IRule% static inline constinit thread_local uint64_t PrivateHapax = 0 ; 
 %\IRule% static inline constinit std::atomic<uint64_t> HapaxAllocator alignas(128) {0} ; 
 %\IRule%  
 public : 
 %\IRule% inline auto %{\fboxsep1pt\colorbox{yellow!50}{operator+}}%(std::invocable auto && csfn) %$\rightarrow$% void {
 %\IRule%  %\IRule%  
 %\IRule%  %\IRule% // Acquire the lock ...
 %\IRule%  %\IRule% // First, conjure a unique hapax nonce for this specific acquire-release 
 %\IRule%  %\IRule% // episode. 
 %\IRule%  %\IRule% // The hapax is globally and temporally unique and specific to 
 %\IRule%  %\IRule% // this thread, this lock, and this acquisition episode
 %\IRule%  %\IRule% auto hapax = PrivateHapax ++ ; 
 %\IRule%  %\IRule% if ((hapax & 0xFFFF) == 0) [[unlikely]] { 
 %\IRule%  %\IRule%  %\IRule% // crossed edge of block allocation 
 %\IRule%  %\IRule%  %\IRule% // current block is exhausted so must reprovision
 %\IRule%  %\IRule%  %\IRule% // In this particular example the high 48-bits of the 64-bit hapax 
 %\IRule%  %\IRule%  %\IRule% // encode the thread "zone" and the lower 16 are the sub-sequence 
 %\IRule%  %\IRule%  %\IRule% // from which the thread can allocate locally.  
 %\IRule%  %\IRule%  %\IRule% hapax = HapaxAllocator.fetch_add(1)+1 ; 
 %\IRule%  %\IRule%  %\IRule% assert (hapax %$\neq$% 0) ; 
 %\IRule%  %\IRule%  %\IRule% hapax = hapax << 16 ; 
 %\IRule%  %\IRule%  %\IRule% assert (hapax > PrivateHapax) ; 
 %\IRule%  %\IRule%  %\IRule% PrivateHapax = hapax + 1 ; 
 %\IRule%  %\IRule% }
 %\IRule%  %\IRule%  
 %\IRule%  %\IRule% assert (hapax %$\neq$% 0) ;       // by convention, 0 is reserved   
 %\IRule%  %\IRule% assert (hapax %$\neq$% Depart.load ()) ; 
 %\IRule%  %\IRule% assert (hapax %$\neq$% Arrive.load ()) ; 
 %\IRule%  %\IRule% auto pred = %\FastPath{Arrive}%.exchange (hapax) ; 
 %\IRule%  %\IRule% assert (pred %$\neq$% hapax) ; 
 %\IRule%  %\IRule%  
 %\IRule%  %\IRule% if (%\FastPath{Depart}%.load(std::memory_order_acquire) %$\neq$% pred) { 
 %\IRule%  %\IRule%  %\IRule% // Contention : slow-path waiting 
 %\IRule%  %\IRule%  %\IRule% // Initial lock acquire can not encounter contention ...
 %\IRule%  %\IRule%  %\IRule% // pred can only be 0 on the initial locking request.
 %\IRule%  %\IRule%  %\IRule% // As we can not have contention on the initial request, 
 %\IRule%  %\IRule%  %\IRule% // we know pred %$\neq$% 0.
 %\IRule%  %\IRule%  %\IRule% assert (pred %$\neq$% 0) ; 
 %\IRule%  %\IRule%  %\IRule% auto slot = ToSlot(pred) ; 
 %\IRule%  %\IRule%  %\IRule%  
 %\IRule%  %\IRule%  %\IRule% // Try to register as a visible waiter ...
 %\IRule%  %\IRule%  %\IRule% if (slot%$\rightarrow$%%\FastPath{VisibleWaiter}%.cas (0, pred) %$\neq$% 0) [[unlikely]] { 
 %\IRule%  %\IRule%  %\IRule%  %\IRule% // The slot is occupied by some other waiter because of a 
 %\IRule%  %\IRule%  %\IRule%  %\IRule% // hash collision 
 %\IRule%  %\IRule%  %\IRule%  %\IRule% // We failed to install ourselves 
 %\IRule%  %\IRule%  %\IRule%  %\IRule% // we expect this situation to be rare
 %\IRule%  %\IRule%  %\IRule%  %\IRule% // Fallback and revert to either Tidex global spinning or 
 %\IRule%  %\IRule%  %\IRule%  %\IRule% // baseline hapax waiting and operate as invisible waiter
 %\IRule%  %\IRule%  %\IRule%  %\IRule% while (%\Global{Depart}%.load (std::memory_order_acquire) %$\neq$% pred) {
 %\IRule%  %\IRule%  %\IRule%  %\IRule%  %\IRule% Pause()  ; 
 %\IRule%  %\IRule%  %\IRule%  %\IRule% }
 %\IRule%  %\IRule%  %\IRule%  %\IRule% goto EnterCS ; 
 %\IRule%  %\IRule%  %\IRule% }
 %\IRule%  %\IRule%  %\IRule%  
 %\IRule%  %\IRule%  %\IRule% // successfully registered and emplaced ourselves as a visible waiter
 %\IRule%  %\IRule%  %\IRule% // We now occupy the slot
 %\IRule%  %\IRule%  %\IRule% // ratify -- Close race window vs thread in unlock() 
 %\IRule%  %\IRule%  %\IRule% // We expect that Depart will remain, with high probability, resident
 %\IRule%  %\IRule%  %\IRule% // in our local cache, given that we just fetched it, above.  
 %\IRule%  %\IRule%  %\IRule% // Under contention, the 1st access, above, is liable to incur a 
 %\IRule%  %\IRule%  %\IRule% // coherence miss, but the access below will usually be satisfied 
 %\IRule%  %\IRule%  %\IRule% // from the cache without a miss. 
 %\IRule%  %\IRule%  %\IRule% if (%\FastPath{Depart}%.load() == pred) [[unlikely]]  { 
 %\IRule%  %\IRule%  %\IRule%  %\IRule% // Inopportune interleaving vs corresponding unlock() 
 %\IRule%  %\IRule%  %\IRule%  %\IRule% // Extricate ourselves from the Waiting array slot
 %\IRule%  %\IRule%  %\IRule%  %\IRule% // we raced vs unlock() and are now the owner -- release the slot
 %\IRule%  %\IRule%  %\IRule%  %\IRule% // We expect this to be rare 
 %\IRule%  %\IRule%  %\IRule%  %\IRule% // Rescind and cancel state as visible waiter 
 %\IRule%  %\IRule%  %\IRule%  %\IRule% // Recant and "unpublish" 
 %\IRule%  %\IRule%  %\IRule%  %\IRule% // Need to use cas(pred,0) as racing unlock() might have already 
 %\IRule%  %\IRule%  %\IRule%  %\IRule% // set the value to 0
 %\IRule%  %\IRule%  %\IRule%  %\IRule%  slot%$\rightarrow$%%\Global{VisibleWaiter}%.cas (pred, 0) ; 
 %\IRule%  %\IRule%  %\IRule%  %\IRule%  goto EnterCS ; 
 %\IRule%  %\IRule%  %\IRule% }
 %\IRule%  %\IRule%  %\IRule%  
 %\IRule%  %\IRule%  %\IRule% // Confirmed -- normal preferred mode of waiting
 %\IRule%  %\IRule%  %\IRule% // The waiting thread is "settled" and emplaced
 %\IRule%  %\IRule%  %\IRule% // Primary busy-wait waiting loop ... 
 %\IRule%  %\IRule%  %\IRule% while (slot%$\rightarrow$%%\HandOver{VisibleWaiter}%.load (std::memory_order_acquire) == pred) {
 %\IRule%  %\IRule%  %\IRule%  %\IRule% Pause() ; 
 %\IRule%  %\IRule%  %\IRule% }
 %\IRule%  %\IRule% } 
 %\IRule%  %\IRule%  
 %\IRule%  %\IRule% EnterCS: 
 %\IRule%  %\IRule% assert (Depart.load() %$\neq$% Arrive.load()) ;
 %\IRule%  %\IRule% // Execute the critical section, expressed as a lambda  
 %\IRule%  %\IRule% // Pass hapax value as context from lock to corresponding unlock
 %\IRule%  %\IRule% csfn() ;
 %\IRule%  %\IRule% assert (Depart.load() %$\neq$% Arrive.load()) ; 
 %\IRule%  %\IRule%  
 %\IRule%  %\IRule% // Unlock() ...
 %\IRule%  %\IRule% assert (Depart.load() %$\neq$% hapax) ; 
 %\IRule%  %\IRule%  
 %\IRule%  %\IRule% // Rendezvous with successor ...
 %\IRule%  %\IRule% // Attempt fast handover to a potential waiter -- successor
 %\IRule%  %\IRule% // inform waiting successor that we are renouncing or 
 %\IRule%  %\IRule% // relinquishing ownership
 %\IRule%  %\IRule% // This is the preferred path for succession and handover 
 %\IRule%  %\IRule% // under contention
 %\IRule%  %\IRule% // CONSIDER: hoist ToSlot(hapax) calculation "up" and outside before CS 
 %\IRule%  %\IRule% auto slot = ToSlot(hapax) ; 
 %\IRule%  %\IRule%  
 %\IRule%  %\IRule% // Consider : Polite LD-if-ST or LD-if-CAS or naked agro CAS() 
 %\IRule%  %\IRule% if (slot%$\rightarrow$%%\HandOver{VisibleWaiter}%.cas (hapax, 0) == hapax) { 
 %\IRule%  %\IRule%  %\IRule% // success!  effected direct handover to specific waiting successor 
 %\IRule%  %\IRule%  %\IRule% // We can safely elide store into Depart() 
 %\IRule%  %\IRule%  %\IRule% // As hapax values are non-recurring, no cleanup to remove values 
 %\IRule%  %\IRule%  %\IRule% // from the array is required
 %\IRule%  %\IRule%  %\IRule% return ; 
 %\IRule%  %\IRule% }
 %\IRule%  %\IRule%  
 %\IRule%  %\IRule% // Cases :
 %\IRule%  %\IRule% // @  no waiters -- uncontended case 
 %\IRule%  %\IRule% // @  waiter but we have hash collisions so our specific successor 
 %\IRule%  %\IRule% //    could not make itself visible
 %\IRule%  %\IRule% //    we expect this to be rare 
 %\IRule%  %\IRule% // @  waiter but our waiter was slow to make itself visible -- racy 
 %\IRule%  %\IRule% %\Global{Depart}%.store (hapax) ; 
 %\IRule%  %\IRule%  
 %\IRule%  %\IRule% // Need to CAS(hapax,0) on the wait array slot to close race window vs 
 %\IRule%  %\IRule% // potential tardy-sluggish arriving waiter.  
 %\IRule%  %\IRule% // Conservatively recover from potential race - compensate
 %\IRule%  %\IRule% // CONSIDER: ST-CAS vs polite ST-LD-if-CAS 
 %\IRule%  %\IRule% slot%$\rightarrow$%%\Global{VisibleWaiter}%.cas (hapax, 0) ; 
 %\IRule% } 
} ;

\end{lstlisting}  

With positive handover, \texttt{unlock} can also skip updating the \texttt{Depart} field during egress.
if it can positively and definitively detect that it accomplished handover to a waiting successor.   
Furthermore, the successor can then also skip checking the \texttt{Depart} field.  
This optimization reduces write invalidations on the shared \texttt{Depart} variable.  
\texttt{unlock} only needs to update the \texttt{Depart} field when there is no contention or
if there were collisions in the waiting array.

Visible waiters completely occupy a slot for the duration of the waiting phase, which means we might be more 
exposed to collisions if we have a large number of waiters.   As such, a given slot can not be shared amongst 
multiple concurrent waiters.   
We note, however, that there is almost no particular downside to increasing the size of the array to statistically
reduce the incidence of collisions.   
And we could, if desired, shift hapax zones if we detect collisions, in an agile and adaptive fashion,
to reduce the odds of future collisions, providing a form of collision avoidance.
Currently a thread falls back to simple Tidex global spinning if a waiter finds the slot occupied.

We also note that a thread could, as an optional optimization,  operate briefly as a visible waiter, 
and if handover was not accomplished after a brief period, to avoid monopolizing an element, 
it could voluntarily vacate its slot and then shift back to the invisible waiting policy in the baseline algorithm.  

The positive handover optimization enabled by visible waiters affords the most relative benefit 
under high flux situations, with high arrival and departure traffic rates on a given lock. 
If threads fail to acquire the lock during the brief initial visible spinning phase, then
we know that throughput and progress over the contented lock is low, 
so the benefit of visible waiting and positive handover will not be as pronounced,
so reverting to invisible waiting does not appreciably degrade performance in this particular
operating regime.  But by shifting to invisible waiting, our threads can reduce load on the visible
waiting array, and allow other threads and locks to benefit from using positive handover via
the waiting array.

Because of the possibility of collisions, the worst-case Remote Memory Reference (RMR)  
complexity \cite{stoc08-attiya,podc02-anderson} for \Hapax{} is unbounded.  
If \emph{all} hapax values were to inadvertently hash to just one singleton slot, then operation 
devolves to the underlying Tidex algorithm, with \emph{global spinning}.

\Invisible{Surrender; abdicate; yield; cede}

\Invisible{phase change for lock from visible to invisible; emergent behavior}  
\Invisible{Altruism; Altruistic; polite; } 
\Invisible{Rescind; Retract; Cancel; Annul; Abandon; Abdicate; revert; vacate; cede; yield; surrender; withdraw; recant; unpublish; }

\section{Related Work}

\Invisible{semi-local; semi-private; quasi-private; 
commonly, frequently, often, usually mostly ::: fere- ; pene- ; quotide ; semi-; cotidie ; vulgo-; plerumque ; saepe- } 

While mutual exclusion remains an active research topic
\cite{ppopp17-Ramalhete,craig-clh,ppopp91-Mellor-Crummey,
EuroPar19-TWA,icdcn20-jayanti,Jayanti-Dissertation,
EuroPar19-GCR,opodis17-dvir,
EuroSys19-CNA,
eurosys17-dice,oopsla99-agesen,topc15-dice,ScottB24,MCSH,isca10-eyerman,aksenov-CLH,aksenov-MCS,tocs19-Guerraoui}  
we focus on locks closely related to our design. 

Simple test-and-set or polite test-and-test-and-set \cite{ScottB24} locks 
are compact and exhibit excellent latency for uncontended operations, but fail 
to scale and may allow unfairness and even indefinite starvation.  Ticket and Tidex locks are compact
and FIFO, and also have excellent latency for uncontended operations, but they also 
fail to scale because of global spinning, although some variations attempt
to overcome this obstacle, at the cost of increased space
\cite{EuroPar19-TWA,Ticket-AWN,dc03-Anderson,tpds90-Anderson,spaa11-dice}.  
For instance, Anderson’s array-based queueing lock \cite{dc03-Anderson,tpds90-Anderson} 
is based on Ticket Locks but provides local spinning.  It employs a waiting 
array for each lock instance, sized to ensure there is at least one array element 
for each potentially waiting thread, yielding a potentially large footprint. 
The maximum number of participating threads must be known in advance when
initializing the lock.  TWA\cite{EuroPar19-TWA} is a variation on ticket locks
that reduces the incidence of global spinning.  

Queue-based locks such as MCS or CLH are FIFO and provide local spinning and
are thus more scalable.  
MCS is used in the linux kernel in the low-level ``qspinlock'' construct
\cite{linux-locks,Long13,LWN2014}.  
Modern extensions of MCS edit the queue
order to make the lock \emph{NUMA-Aware}\cite{EuroSys19-CNA}. 
MCS readily allows editing and re-ordering of the queue of waiting threads,
\cite{markatos,eurosys17-dice,EuroSys19-CNA} whereas editing the
chain is more difficult under CLH, HemLock and \recipro{}\cite{arxiv-recipro,ppopp25-dice}, where
there are no explicit linked lists. 

CLH is extremely simple, has excellent Remote Memory Reference (RMR) 
complexity\cite{stoc08-attiya,podc02-anderson}, and requires just a
single atomic exchange operation in the \Acquire{} operation and no atomic 
read-modify-write instructions in \Release{}. 
Unfortunately the waiting elements migrate between threads, which may be
inimical to performance in NUMA environments.  CLH locks also require 
explicit constructors and destructors, which may be inconvenient or preclude their use in some environments
\footnote{ 
Many lock implementations require just trivial constructors to set the lock fields to 0 or some constant,
and trivial destructors, which do nothing.  The Linux kernel \texttt{spinlock\_t}/\texttt{qspinlock\_t}
interface provides a constructor, but does not even expose a destructor. Such decisions
influence and may limit which lock algorithms can be readily implemented under a particular interface.}. 
Similarly, C++ \texttt{std::mutex} is allowed to be \emph{trivially destructible}, meaning
storage occupied by trivially destructible objects may be reused without calling the destructor.  
Under both GCC g++ version 15 and Clang++ Version 20, \texttt{is\_trivially\_destructible} reports \texttt{True}
for \texttt{std::mutex}, and as such, destructors do not run. 

Our specific CLH implementation uses a variation on Scott's \cite{ScottB24} Figure 4.14, which
converts the CLH lock to be \emph{context-free}\cite{ppopp16-wang}, adhering to a simple 
programming interface that passes just the address of the lock, albeit at the cost of
adding an extra field to the lock body to convey the address of the
head waiting element to the corresponding \Release{}.  This field is protected by the lock itself. 


The K42 \cite{K42,ScottB24} variation of MCS can recover the queue element before returning from \Acquire{}
whereas classic MCS recovers the queue element in \Release{}.  
That is, under K42, a queue element is needed only while waiting but not while the lock is held,
and as such, queue elements can always be allocated on stack, if desired. 
While appealing, the paths are much more complex and touch more cache lines than the classic
version, impacting performance.  In addition, neither the K42 doorway nor the \Release{} path operate
in constant time.

HemLock combines aspects of both CLH and MCS to form a lock
that has very simple waiting node memory lifecycle constraints, is \emph{context-free} but
still scales well in common usage scenarios.  
HemLock requires just a singleton waiting element (queue node) per thread, which
can be placed in thread-local storage. 
HemLock does not provide constant remote memory reference (RMR) complexity 
\cite{opodis17-dvir} in scenarios where a thread holds multiple contended locks. 
In this situation the single node suffers from \emph{multi-waiting}\cite{spaa21-Dice}.  
Similar to MCS, HemLock lacks a constant-time unlock operation, whereas
the unlock operator for CLH and Ticket locks is constant-time.  Unlike MCS, HemLock requires
active synchronous back-and-forth communication in the unlock path between the outgoing
thread and its successor to protect the lifecycle of the waiting element. 
We note, however, that HemLock remains constant-time in the \Release{} operator to the
point where ownership is conveyed to the successor.  
HemLock uses \emph{address-based} transfer of ownership, writing the address
of the lock instead of a boolean, differentiating it from MCS and CLH. 
\recipro{}\cite{arxiv-recipro,ppopp25-dice}, like HemLock, requires just a singleton 
per-thread waiting element allocated in thread-local storage.  
\recipro{} has constant-time arrival and unlock paths, but is not FIFO, 
instead providing 2-bounded-bypass.  




We opted to exclude NUMA-aware locks such as Cohort Locks \cite{topc15-dice,PPoPP12-dice} and 
Compact NUMA-Aware Locks (CNA) \cite{EuroSys19-CNA} from consideration.
We excluded Fissile Locks and GCR\cite{EuroPar19-GCR} as they have 
lax time-based anti-starvation mechanism.
Fissile, specifically, depends on the owner of the \emph{inner lock} to make progress
to monitor for starvation.  

We also excluded MCSH\cite{MCSH} which is a recent variation of MCS
that uses on-stack allocation of queue nodes and hence supports a standard locking interface.  
Like MCS, the lock is not innately context-free, and additional information needs to
be passed from \Acquire{} to \Release{} via extra fields in the lock body.  
In our experiments, the performance of MCSH is typically on par with MCS proper.

\Invisible{We excluded MCSH\cite{MCSH}, which is essentially a simplified Fissile lock that uses MCS instead 
of MCS-based NUMA-aware CNA for the inner lock, removes the fast-path \emph{barging} 
attempt on arrival, has \emph{patience} set to 0, yielding FIFO admission, and uses 
Fissile's ''partial MCS release'' optimization to defer the final step in releasing the 
inner lock until the \Release{} operation.} 

We further restrict our comparison to locks that use \emph{direct succession} and 
hand off ownership directly from the owner to a specific successor and
do not admit \emph{barging} or \emph{pouncing} where the lock is released even though
waiting threads exist, which allow newly arriving threads to seize the lock.

\Invisible{Seize; snarf; grab; raptor; barge; usurp; arrogate; pounce; } 


\Invisible{Exclude locks that use \emph{barging}}

\newcommand{\NOTEA}{\hyperlink{HemLock}{\blu{Note-1}}}

\medmuskip=0mu\relax
\thinmuskip=0mu\relax
\thickmuskip=0mu\relax


\begin{landscape} 
\begin{table*} [ht!]
\scriptsize 
\centering
\begin{tabular}{llllllllll}
\toprule

\multicolumn{1}{l}{} &
\multicolumn{9}{c}{Lock Algorithm}      \\
\cmidrule(lr){2-10}

\multicolumn{1}{l}{Property} &
\multicolumn{1}{l}{\rot{MCS}} &
\multicolumn{1}{l}{\rot{CLH}} &
\multicolumn{1}{l}{\rot{HemLock}} &
\multicolumn{1}{l}{\rot{Ticket}} &
\multicolumn{1}{l}{\rot{TWA}} &
\multicolumn{1}{l}{\rot{Recipro}} &
\multicolumn{1}{l}{\rot{Tidex}} &
\multicolumn{1}{l}{\rot{Hapax}} &
\multicolumn{1}{l}{\rot{HapaxVW}} \\
\midrule
Spinning                             & L+P       & U+P       & L+S    & \red{U+G}      & U+S               & L+P      & \red{U+G}     & U+S    & U+S      \\
Constant-time \Release{}             & No        & Yes       & No$*$  & Yes            & Yes               & Yes      & Yes           & Yes    & Yes      \\
Context-free                         & No        & No        & Yes    & Yes            & Yes               & No       & No            & No     & No       \\ 
FIFO                                 & Yes       & Yes       & Yes    & Yes            & Yes               & \red{No} & Yes           & Yes    & Yes      \\
Path Complexity -- Branches          & 2+3       & 2+0       & 2+2    & 1+0            & 4+0               & 3+2      & 1+0           & 4+0    & 6+1      \\
On-Stack                             & No        & No        & No     & N/A            & N/A               & Possible & N/A           & N/A    & N/A      \\ 
Nodes Circulate                      & No        & \red{Yes} & No     & N/A            & N/A               & No       & N/A           & N/A    & N/A      \\
Explicit CTOR/DTOR Required          & No        & \red{Yes} & No     & No             & No                & No       & No            & No     & No       \\
Thread CTOR/DTOR Required            & \red{Yes$*$} & \red{Yes} & No     & No          & No                & No       & No            & No     & No       \\
Max Remote Misses per episode        & 4         & 4         & 4      & \red{$T$}      & \red{$T\cdot{}2$} & 2        & \red{$T$}     & 4      & 4        \\
Invalidations per episode            & 6         & 5         & 5      & \red{10($T$)}  & \red{8.5($T$)}    & 4        & \red{10($T$)} & 5      & 4        \\
Dynamic Memory Allocation            & \red{Yes} & \red{Yes} & No     & No             & No                & No       & No            & No     & No        \\
Trivial TryLock                      & Yes       & \red{No}  & Yes    & 64-bit         & 64-bit            & Yes      & \red{No}      & Yes    & Yes      \\
Space Complexity                     & $(S\cdot{}L)+(E\cdot{}A)$ & $(S\cdot{}L) + (E\cdot{}(L+T))$ & $L + (E\cdot{}T)$ & $S\cdot{}L$  & $(S\cdot{}L) + 4096$  & $(S\cdot{}L) + (E\cdot{}T)$ & $S\cdot{}L$ & $(S\cdot{}L) + 4096$ & $(S\cdot{}L) + 4096$ \\

\midrule[\heavyrulewidth]
\bottomrule
\end{tabular}%
\vspace{1cm} 
\caption{Comparison of lock algorithm properties}\label{Table:Compare}
\end{table*}
\end{landscape}

In Table-\ref{Table:Compare} we compare the attributes of various lock algorithms.  
\red{Red}-colored cells indicate potential undesirable properties.  
Note that all the locks provide a constant-time doorway phase.
In the following we explain the meaning of each property.   


\BoldSection{Spinning}: \textbf{L} indicates that waiters spin on a location that is usually collocated 
in the system topology with the waiting thread (typically on-stack or in thread-local storage) , which may confer performance benefits on 
some systems, such as those that use the Intel UPI\cite{UPI} \emph{home-based} coherence protocol, while
\textbf{U} indicates that the location being waited upon is generally ``unknown'' and may be either local or remote to
the waiting thread.  \textbf{P} indicates \emph{private} spinning, where at most
one thread spins on a given location at a given time, \textbf{G} indicates global spinning,
where all threads waiting on a given lock busy-wait on a single location, and \textbf{S} 
indicate \emph{semi-local}\cite{spaa21-Dice} spinning, where threads, absent hash collisions,
enjoy private spinning.

\BoldSection{Constant-time Unlock}: indicates if the lock \Release{} operation is bounded.  
As noted above, HemLock is constant-time up to the point where the lock is either released or transferred to a successor, 
but the \Release{} operator, for reasons of memory safety, then waits for the successor to 
acknowledge transfer of ownership before the memory underlying the queue element can be potentially reused.
Specifically, in HemLock an uncontended \Release{} operation is constant-time and a contended \Release{} is 
constant-time up to and including the point where ownership is conveyed to the successor. 

\BoldSection{FIFO}: indicates the lock provides strict FIFO admission.  The only non-FIFO lock
is \recipro{}, which still provides 2-bounded-bypass and starvation avoidance. 

\BoldSection{Context-free}: indicates additional information does not need to be transferred from the \Acquire{}
operator to the corresponding \Release{} operation.

\BoldSection{Path Complexity - branches}: We tally the number of conditional branches in the \Acquire{} and
\Release{} methods, respectively, using the 
platform independent LLVM intermediate representation (IR) instructions, as emitted by \texttt{clang++-20},
which serves as a simple measure for path complexity.  
All \Acquire{} operators have at least one branch, by virtue of the waiting loop. 
In some cases, we observed that the compiler converts \texttt{if} statements to conditional move operations, 
shifting control flow to data flow.  

\BoldSection{On-Stack}: indicates the queue elements, if any, may be allocated on-stack.  This also
implies the nodes do not migrate and have a tenure constrained to the duration of the locking episode. 

\BoldSection{Nodes Circulate}: queue elements migrate between threads.  This often implies the need for
an explicit queue element lifecycle management system and precludes convenient on-stack allocation
of queue elements.  Migration may also be unfriendly to performance in NUMA environments.  

\BoldSection{Explicit CTOR/DTOR Required}: indicates the lock requires non-trivial constructors or destructors.  
CLH, when used in general purpose environments, for instance, requires destructors to run to release the wait elements referenced in the lock, to
avoid memory leaks.   

\BoldSection{Thread CTOR/DTOR Required}: indicates if per-thread constructors or destructors are needed to 
initialize or cleanup after dynamically created threads.  We note that MCS implementations, could, if 
necessary, allocate queue nodes as needed, dynamically, in \Acquire{}, and then release such queue nodes in
\Release{}.  This avoids the need for thread-level destructors.  
(To avoid circular or recursive dependencies, we further require a specialized node allocator that
is implemented via different types of locks).  To avoid the allocator becoming a contention hotspot and bottleneck,
common MCS implementations use per-thread caches of available elements, which need to be released when 
the thread terminates.  Similarly, CLH could, in theory, allocate and free queue nodes on demand, although
most implementations use a thread-local cache containing a singleton pointer to a free queue node,
as seen in Scott's \cite{ScottB24} Figure 4.14.  
MCSH\cite{MCSH} and K42\cite{K42,ScottB24} MCS variations side step this concern by allocating elements on-stack,
but at the cost of added complexity.  

\BoldSection{Max Remote Misses per episode}: is the worst-case maximum number of misses to \emph{remote} memory
incurred, under simple sustained contention, by a matching \Acquire{}-\Release{} pair.  Misses to \emph{remote} memory --
memory \emph{homed} on a different NUMA node --
may be more expensive than local misses on various platforms, such as modern Intel systems that use 
the UPI coherence fabric \cite{UPI}, where miss requests are first adjudicated by the \emph{home node} of the cache line. 
Algorithms where nodes circulate are more vulnerable to accumulating such remote misses.  
For HemLock, we assume simple contention with no \emph{multi-waiting}.  
We derived the ``Max Remote Misses per episode'' value via static inspection of the code, 
and determining which of the accesses that might cause coherence misses might also be to locations 
homed on remote nodes.  We assume that back-to-back accesses of shared variables that reside on the same cache
line, by a given thread,  will result in just one miss.  And for Hapax Locks we assume that the hapax value does not 
need to be replenished, which would incur an addition likely remote miss for the \texttt{fetch-and-add} atomic
operation.  We further assume no collisions (or near collisions, which could induce false sharing) in the waiting array.  

\BoldSection{Invalidations per episode}:  is the number of coherence misses, under sustained contention,
experienced by an \Acquire{}-\Release{} episode in a given thread.   
We empirically approximate that number as follows.  
Using an ARMv8 system we modified the MutexBench microbenchmark (described below) to have a degenerate critical section that advanced
only a \emph{thread local} random number generator and to pass any context from \Acquire{} to \Release{} via 
thread-local storage, in order to reduce mutation of shared memory.  Absent coherence misses, an \Acquire{}-\Release{} episode,
including the critical section, manages to remain completely resident in the private L1 data cache. As such, any misses
are coherence misses.  
This technique yields a useful empirical metric on how much coherence traffic various lock algorithms generate.
We used the ARM \texttt{l2d\_cache\_inval} hardware performance counter, which 
tallies L2 cache invalidation events, as simple proxy for coherence traffic latency and bandwidth.  
We report the number of \texttt{l2d\_cache\_inval} events per episode. As can be seen, \recipro{} and \Hapax{} with
visible waiters are parsimonious and incur just 4 invalidations per episode, while CLH requires 5
\footnote{We observed similar ratios with performance counters that reported the number of ``snoop'' messages
on the ARM coherent mesh ``CMN'' device,
and on Intel by counting \texttt{OFFCORE} requests.
We also observed in passing, while examining data from hardware performance counters, that CLH and MCS suffered 
from more stalls (both event counts and duration) from misses than did Reciprocating locks.  
CLH in particular executes a dependent load in the critical path in the arrival doorway on the address returned 
from the atomic \texttt{exchange}. 
The address to be loaded from is not known until after the \texttt{exchange} returns,
denying the processor the ability to speculate or execute out-of-order.}.  
Ticket locks and Tidex require $T$ (10 in our example) invalidations per episode, where $T$ is the number of participating threads, 
given the global spinning.  The number of misses incurred by CLH, MCS, HemLock, Reciprocating Locks and \Hapax{} (assuming
no hash collisions) is constant and not a function of the number of threads. 
These empirically-derived results align closely with a static analysis of the code and the expected number of 
coherence misses in the \Acquire{} and \Release{} paths.  
As the coherent interconnect fabric is a shared resource, excessive coherence traffic and 
invalidation rates can throttle scalability, and even impact the performance of other unrelated threads running on the system.  

\Invisible{parsimonious; frugal; economical;} 

\BoldSection{Dynamic Memory Allocation}: indicates if the algorithm requires some type 
of dynamic memory allocation, typically for waiting elements. 
We note that MCS can avoid allocation and use on-stack allocation if all locking is lexically balanced and
that CLH can also avoid allocation -- by initially pairing elements 1:1 with locks and threads -- 
if all threads and locks are statically allocated in advance of execution or where dynamically allocated threads and locks are effectively immortal.  
These advantageous circumstances are not normally found or available in general-purposes environments.  

\BoldSection{Trivial TryLock}: indicates if the algorithm admits a trivial wait-free \texttt{trylock} operator.
We note that Ticket-based locks can implement \texttt{trylock} if the \texttt{Grant} and \texttt{Ticket} fields 
are 64-bit bits, and effectively immune from roll-over overflow aliasing\footnote{  
Let’s say our ticket lock implementation has 32-bit \texttt{Ticket} and \texttt{Grant} fields. 
The naive but flawed way to implement \texttt{trylock} is to fetch the \texttt{Grant} field into a temporary $G$,
and then try to CAS \texttt{Ticket} from $G$ to $G+1$.  If execution is delayed between the load and 
the subsequent CAS,  the values could have rolled around and we could suffer a false positive CAS success.   
This situation would be extremely rare, but not impossible.   To avoid exclusion failure, if the CAS was successful, 
we might then try to adapt and remediate the algorithm to fetch and re-check \texttt{Grant} -- after the successful 
CAS -- and if \texttt{Grant} had changed, then we need to wait.  
This approach restricts us to 4G-1 possible concurrently waiting threads, but that’s arguably reasonable.
But our \texttt{trylock} implementation is now flawed in the sense that it might need to wait, to recover.  
Another work-around for 32-bit fields is to collocate the \texttt{Ticket} and \texttt{Grant} fields
within an atomic 64-bit superword and then to use a 64-bit CAS accordingly.  This forces placement of the 2 fields,
and can preclude attempts to limit false sharing by sequestering or isolating the constituent 
fields onto distinct cache lines.  
With 64-bit fields, however, given that rollover is not a practical concern, we can simply
fetch \texttt{Grant} into $G$ and then try to CAS \texttt{Ticket} from $G$ to $G+1$.  
The \texttt{trylock} attempt is successful if and only if the CAS was successful.}.  

\BoldSection{Space Complexity}: reflects the space complexity of the lock algorithm where $T$ is the number of active threads,
$L$ is the number of currently extant locks, $A$ is the number of locks currently held plus the number
of threads currently waiting on locks, $E$ is the size of the waiting element. 
$S$ is the size of a lock instance.  When a lock algorithm
requires context to be passed from \Acquire{} to the corresponding \Release{}, we set $S$ to $2$
to indicate that our implementation allocated an extra word in the lock body to pass such information. 
Implementations that require context can avoid that space requirement in the lock body if they 
opt to pass context information by other means, such as via thread-local storage, in which case $S$ would be $1$.  
HemLock is context-free, so the per-lock space usage is just $1$, while
Tidex, Ticket Locks, TWA and \Hapax{} require 2 words per lock.  
\Hapax{} and TWA require an additional 4096 words for the global waiting table that is shared over all thread and lock instances. 
For MCS and CLH we assume that the implementation stores the head of the chain -- reflecting the current 
owner --  in an additional field in the lock body, and thus the lock consists of 
\texttt{head} and \texttt{tail} fields, requiring 2 words in total.

\Invisible{semi-local; commonly, frequently, often, usually mostly ::: fere- ; pene- ; quotide ; semi-; cotidie ; vulgo-; plerumque ; saepe- }


\section{Empirical Performance Results}

We collected data on 2 different platforms.  
\textbf{ARMv8} is an ARMv8 AARCH64 Ampere Altra Max NeoVerse-N1 non-NUMA system with 128 processors on a single socket,
running Ubuntu 24.04.
We compiled all code using the \texttt{-mno-\allowbreak{}outline-atomics} \texttt{-march=\allowbreak{}armv8.2-a+lse} flags
in order to allow direct use of modern atomic \texttt{exchange}, \texttt{CAS} and \texttt{fetch-and-add} instructions instead
of the legacy LL-SC (load-locked store-conditional) forms thereof.
All lock busy-wait loops used the ARMv8 \texttt{YIELD} instruction.  In the listings, the \texttt{Pause} operation
reflects a single ARMv8 \texttt{YIELD} instruction.  
\textbf{AMD} is a 2-node system with AMD EPYC 9754 128-Core Processors (Ryzen) running Ubuntu 24.04.  
Each processor has 2 hyperthreads, yielding a total of 512 logical processors.  
The \texttt{Pause} operation is implemented as the \texttt{PAUSE} instruction.  

Factory-provided system defaults were used in all cases.  
Unless otherwise noted, default free-range unbound threads were used, with no pinning of threads to processors or NUMA nodes.  
We used the default version of the GNU \texttt{g++} compiler as provided by the Ubuntu distributions.  


We compared \textbf{MCS, CLH, TWA, Tidex, Ticket Locks, HemLock},
Reciprocating Locks(\textbf{Recipro}), Hapax (\textbf{HapaxIW}), 
and Hapax augmented with the visible waiters optimization (\textbf{HapaxVW}). 

\subsection{\texttt{std::atomic<>::exchange()}}  

\begin{figure}[h]                                                                    
\includegraphics[width=7cm]{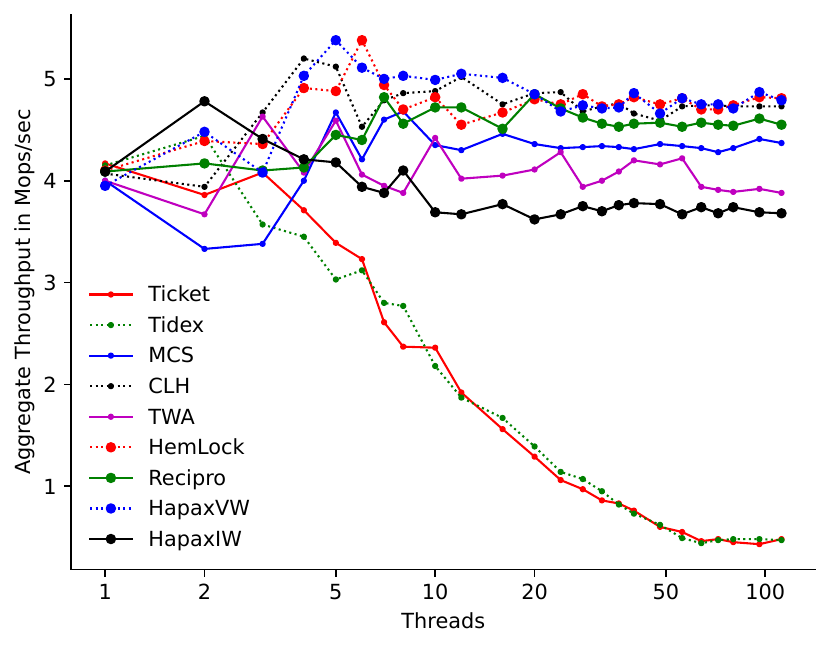}
\vspace{-0.2cm} 
\caption{std::atomic<>::exchange} 
\label{Figure:atomic-exchange-NCSL-100:armv8}
\end{figure}

In Figure-\ref{Figure:atomic-exchange-NCSL-100:armv8} we used a C++ 
benchmark, running on the ARMv8 system, 
that defines a simple structure type \texttt{S} that contains 5 32-bit \texttt{int} fields.  
The benchmark spawns $T$ concurrent threads, each of which loops, 
calling \texttt{std::atomic<S>::exchange()} to swap a local copy with a singleton global instance, 
followed by a non-critical section phase.  
We report the aggregate throughput rate, in completed exchange operations per second, 
at the end of a 10 second measurement interval, and we plot the median of 11 distinct runs in the figure.
The number of threads $T$ is on the X-axis and aggregate throughput in the Y-axis.
The C++ compiler and runtime implement \texttt{std::atomic} for such ``large'' objects by
hashing the address of the instance into an array of pthread mutexes, and acquiring those 
as needed to implement the desired atomic action.  
Concurrent \texttt{exchange} operations on the same instance can thus result in 
contention on those underlying mutexes. 
We used the \texttt{LD\_PRELOAD} facility
to interpose on the pthread locking operations, replacing them with the algorithms of interest.
Interposition allows us to change the underlying mutex without changing the application binary,
by simply changing the \texttt{LD\_PRELOAD} environment variable to reference various
lock implementation libraries, which are loaded at process startup time.  This allows for a fair
comparision, but, like all pthreads implementations, precludes any compile-time optimizations
that might otherwise be enabled at locking sites, between the application code and the 
locking algorithm.  
Any required lock context is collocated adjacent to the fields in the lock body.  

In the non-critical phase, threads invoke a local 
\emph{xoroshiro128plus}\cite{Vigna-xoroshiro-PRNG} pseudo-random number generator (PRNG) to 
generate an integer uniformly distributed in [0,100) and then advance that same PRNG that many steps.  

\renewcommand{\topfraction}{0.85}
\renewcommand{\bottomfraction}{0.85}
\renewcommand{\textfraction}{0.15}
\renewcommand{\floatpagefraction}{0.8}
\renewcommand{\textfraction}{0.1}
\setlength{\floatsep}{5pt plus 2pt minus 2pt}
\setlength{\textfloatsep}{5pt plus 2pt minus 2pt}
\setlength{\intextsep}{5pt plus 2pt minus 2pt}
\setlength{\abovecaptionskip}{1pt} 
\setlength{\belowcaptionskip}{1pt} 

\vspace{-0.2cm} 

\subsection{\texttt{LevelDB}}   

\begin{figure}[h]                                                                    
\includegraphics[width=7cm]{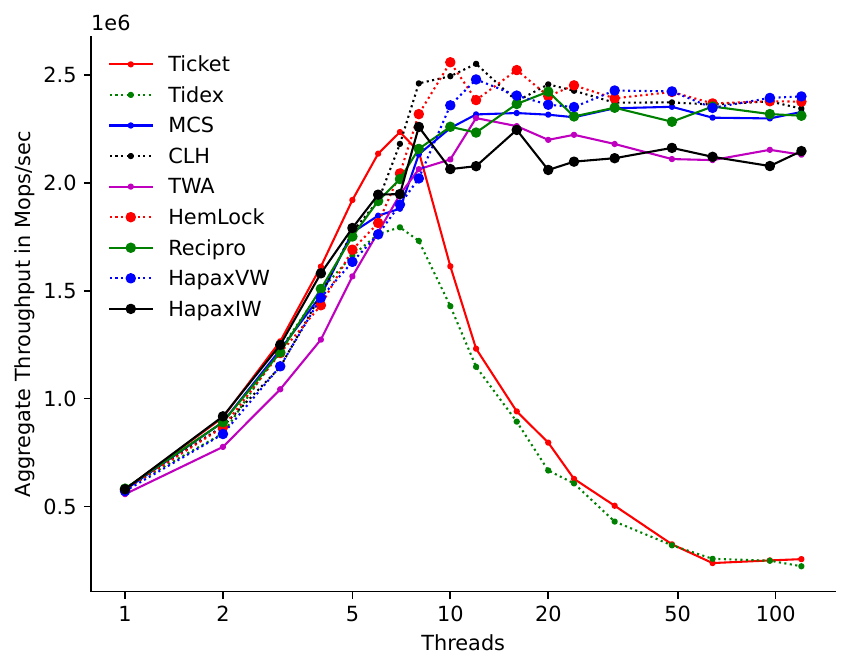}
\vspace{-0.2cm} 
\caption{LevelDB} 
\label{Figure:LevelDB}
\end{figure}

In Figure-\ref{Figure:LevelDB}  we used the ``readrandom'' benchmark in LevelDB version 1.23
database\footnote{\url{leveldb.org}}, again on the ARMv8 system, varying the number of 
threads and reporting throughput
from the median of 5 runs of 50 seconds each.
Each thread loops, generating random keys and then tries to read the associated value from
the database.
We first populated a database\footnote{db\_bench \---\---threads=1
\textendash{}\textendash{}benchmarks=fillseq \---\---db=/tmp/db/}
and then collected data\footnote{db\_bench \---\---threads=\emph{threads}
\---\---benchmarks=readrandom \\ \---\---use\_existing\_db=1
\---\---db=/tmp/db/ \---\---duration=50}.
We made a slight modification to the \texttt{db\_bench} benchmarking
harness to allow runs with a fixed duration that reported aggregate throughput.
LevelDB uses coarse-grained locking, protecting the database with a single central mutex:
\texttt{DBImpl::Mutex}.  Profiling indicates contention on that lock via \texttt{leveldb::DBImpl::Get()}.
The results in Figure-\ref{Figure:LevelDB} largely echo the moderate contention results in Figure-\ref{fig:MutexBench}.

\vspace{-0.2cm} 

\subsection{MutexBench benchmark} 
\label{Section:MutexBench} 

We also used a simple locking microbenchmark, MutexBench, to evaluate the performance of Hapax Locks. 
The MutexBench benchmark spawns $T$ concurrent threads. Each thread loops as follows:
acquire a central lock $L$; execute a critical section; release $L$; execute
a non-critical section. 

For the \emph{moderate contention} configuration, the critical section advances a
shared global 
\emph{xoroshiro128plus}\cite{Vigna-xoroshiro-PRNG} pseudo-random number generator (PRNG) 
one step, and the non-critical section, as configured command-line arguments for this benchmark, 
advances a thread-local PRNG by 500 steps.  
In the \emph{maximum contention} configuration, we set the non-critical section steps to be 0,
yielding an empty non-critical section, subjecting the lock to extreme contention.  
At just one thread, this configuration also constitutes a useful benchmark for uncontended latency.  
At the end of a 10 second measurement interval the benchmark
reports the total number of aggregate iterations of the outer loop (\texttt{lock}-\texttt{unlock} pairs) 
completed by all the threads.

We also report a trivial measure of long-term fairness, expressed as the number of iterations completed 
by the thread that made the least progress in the interval divided by the number of iterations 
completed by the thread that made the most progress.  
A value of $1$ is ideally fair and values toward $0$ less so.  

After the measurement interval, the benchmark harness resets the global shared PRNG state and then
advances, via sequential execution, the PRNG by the tally of steps taken over all the threads.  
As a cheap and racy test of exclusion and safety, we then ensure the shared PRNG re-arrives at the same
state as observed at the end of the interval.  

\begin{figure*}[t!]
\subfloat[Maximum Contention ARMv8]{%
    \includegraphics[width=7cm]{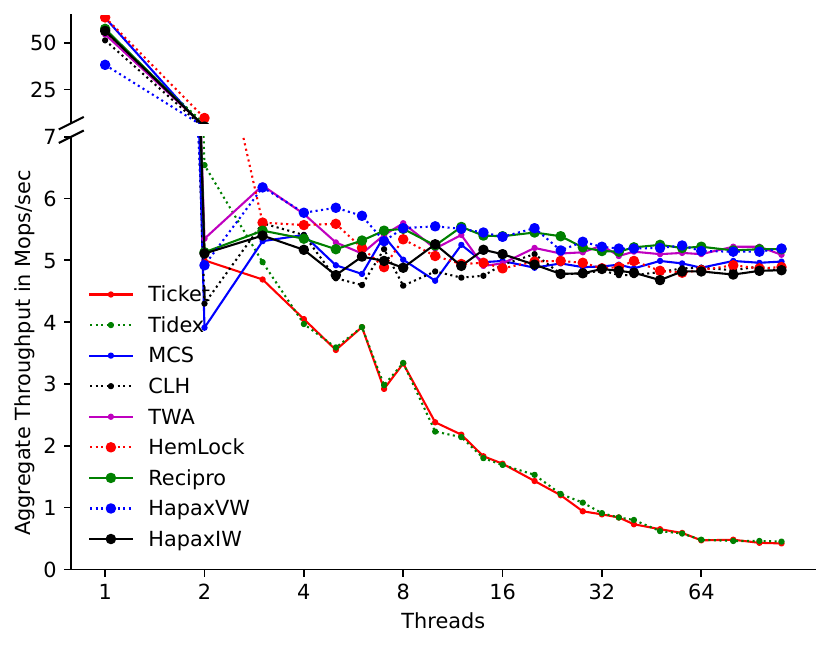}
    \label{Figure:MaximumContention:armv8}
}\hfill
\subfloat[Moderate Contention ARMv8]{%
    \includegraphics[width=7cm]{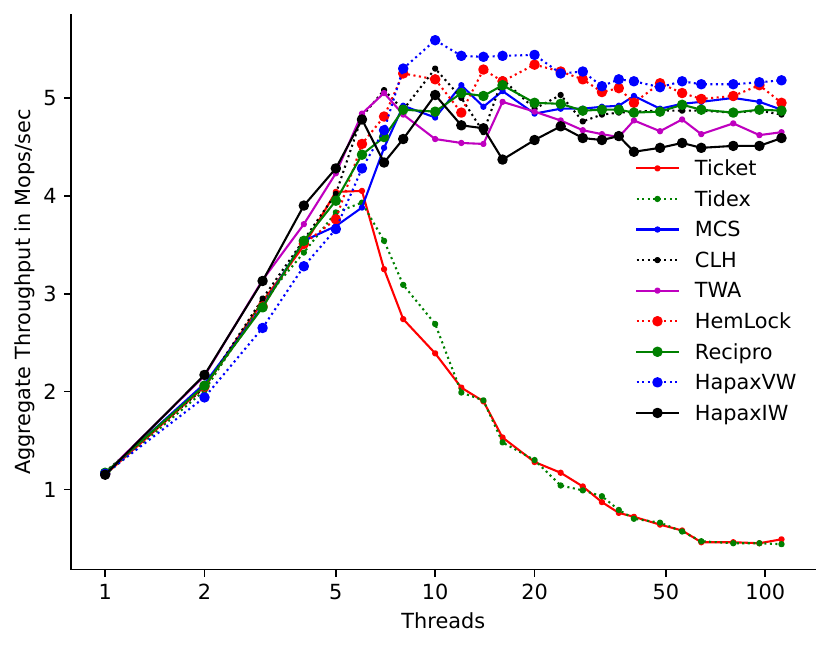}
    \label{Figure:ModerateContention-armv8}
}\\
\subfloat[Maximum Contention AMD]{%
    \includegraphics[width=7cm]{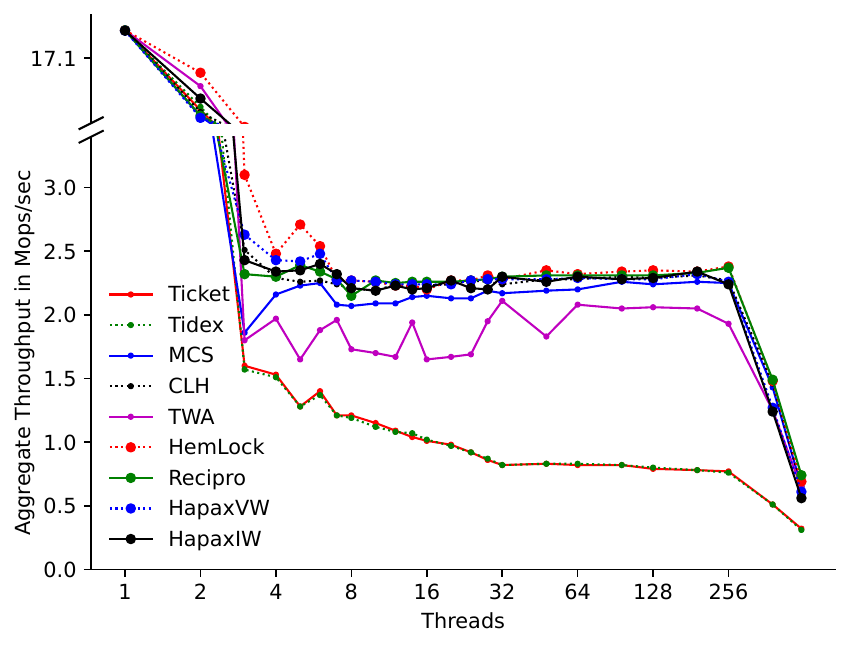}
    \label{Figure:MaximumContention:AMD}
}\hfill
\subfloat[Moderate Contention AMD]{%
    \includegraphics[width=7cm]{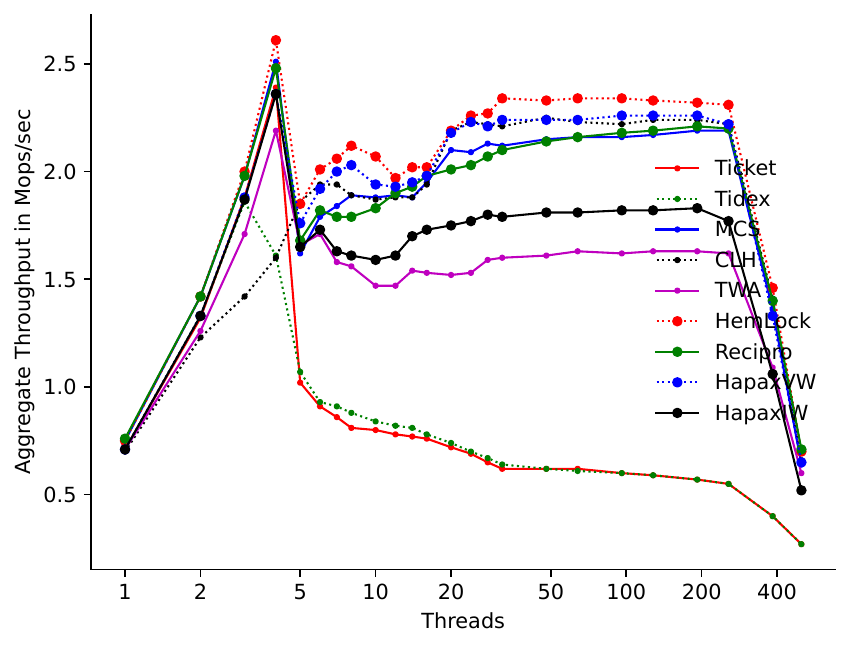}
    \label{Figure:ModerateContention:AMD}
}\\
\subfloat[Maximum Contention AMD-N1]{%
    \includegraphics[width=7cm]{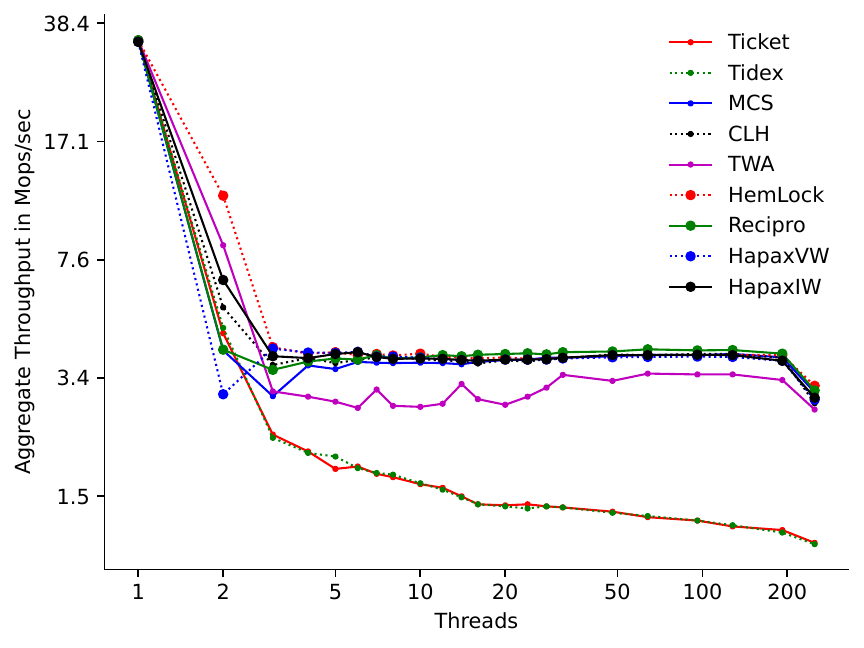}
    \label{Figure:MaximumContention:AMD-N1}
}\hfill
\subfloat[Moderate Contention AMD-N1]{%
    \includegraphics[width=7cm]{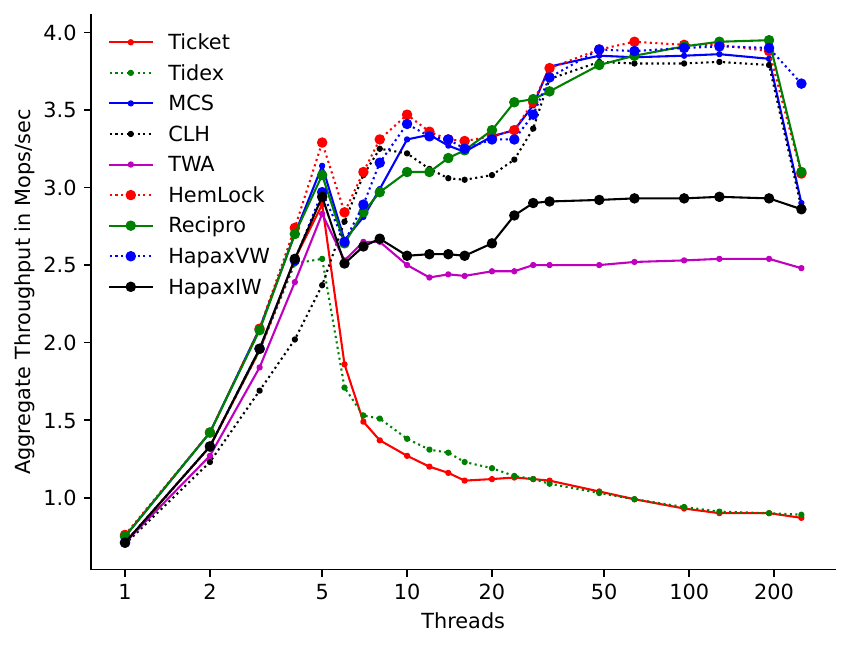}
    \label{Figure:ModerateContention:AMD-N1}
}
\caption{MutexBench}
\label{fig:MutexBench}
\end{figure*}

We repeated each data point 11 times, with independent runs, and reported the median throughput in 
Figure-\ref{fig:MutexBench}.
For clarity and to convey the maximum amount of information to allow a comparison of the algorithms,
the throughput on the $Y$-axis is offset from 0, and, as needed, the $Y$-axis is broken or log scale.  

We note that HapaxVW Locks offer performance on-par with MCS, CLH, TWA, and HemLock, while performance
falls either on-par or slightly under that of Reciprocating Locks.  
Tidex and Ticket locks fail to scale because of global spinning.  

\textbf{AMD-N1} reflects a configuration of the AMD system where constrained execution to just
one NUMA node, with 256 processors (instead of the full complement of 512), in order to better examine NUMA 
performance effects\footnote{numactl --cpunodebind 1 --membind 1}.

\renewcommand{\topfraction}{0.85}
\renewcommand{\bottomfraction}{0.85}
\renewcommand{\textfraction}{0.15}
\renewcommand{\floatpagefraction}{0.8}
\renewcommand{\textfraction}{0.1}
\setlength{\floatsep}{5pt plus 2pt minus 2pt}
\setlength{\textfloatsep}{5pt plus 2pt minus 2pt}
\setlength{\intextsep}{5pt plus 2pt minus 2pt}
\setlength{\abovecaptionskip}{1pt}
\setlength{\belowcaptionskip}{0pt}

\section{Conclusion} 

\Invisible{claim, novel, novelty, contribution} 
\Invisible{Lesson, Informs, Instructs, Guides} 
\Invisible{Exemplar; explainer} 
\Invisible{MRAT = most recently arrived thread} 
\Invisible{strikes compromise; trade-off; balance; tension}  
\Invisible{derangement, disorder} 
\Invisible{Pick random starting position and direction}  
\Invisible{incur; suffer; induce; generate; require} 

\Invisible{KPI}

Unlike CLH, and like TWA, no pointers shift thread ownership in Hapax Locks (using the sense of \emph{ownership} 
employed by the Rust and C++ programming languages \cite{ownership}, where responsibility for disposal
of the referent transfers with the pointer), 
which is appealing as no memory lifecycle issues can exist, making it easier to reason about memory-safety.  
In HemLock, Reciprocating Locks, MCS and MCSH, threads 
spin privately and locally on memory that remains owned by the waiting thread, but the address of that memory
escapes, and is exposed to other threads, which can read and write that memory.   
Under Hapax Locks, no thread-local (\emph{thread owned}) addresses
leak, escape, and are exposed and shared for access by other threads. 

Hapax Locks, being \emph{value-based} instead of address-based, avoid the lifecycle concerns
present in CLH and MCS locks, as those lock algorithms must manage \emph{queue nodes}. 
This property makes Hapax Locks particularly easy to integrate into existing systems,
and in particular for low-level system software where the lock subsystem may not have 
access to facilities such as dynamic memory allocation.  (Dynamic memory allocation facilities, such
as \texttt{malloc}, are commonly implemented using locks, presenting a circular dependency.
CLH and MCS typically allocate queue nodes via dynamic allocation, although the K42\cite{K42} and MCSH variants
are able to allocate nodes on-stack, and those nodes are required to be live only during the acquisition and 
waiting phases and not over the critical section or unlock operation.  Hapax avoids such dependences and entanglements).  
Hapax Locks impose very few requirements or dependencies on the ambient runtime environment. 
Unlike CLH,  Hapax Lock instances require only trivial constructors and destructors,
and unlike both CLH and MCS, no allocation of queue nodes is required.  
As they never require allocation, or calls to any other services,  the 
Hapax \texttt{lock} and \texttt{unlock} methods can be implemented as efficient \emph{leaf} routines.  
And unlike Reciprocating Locks, Hapax Locks provide strict FIFO admission order.

Hapax Locks provide performance on par with existing state of the art algorithms,
such a MCS and CLH.  As it is value-based instead of address-based, as is the case 
for MCS and CLH, it avoids ensuant complications associated with allocating or
deallocating memory for waiting elements, and concerns about the lifecycles thereof.  
The key benefit conferred by Hapax Locks is that they are easy to integrate or retrofit
into existing systems, with minimal requirements, affording additional options for 
lock algorithm selection to systems developers.


\acknowledgements{We thank Peter Buhr at the University of Waterloo for access to ARMv8 and AMD systems.  
We also thank the anonymous PPoPP reviewers for their helpful comments.} 

\bibliography{NonceLock}

\newpage
\appendix
\onecolumn

\section{Key Properties of Hapax Locks} 

\Texticle{}Strict FIFO-FCFS admission order.

\Invisible{FIFE = First-in-first-enabled} 

\Texticle{}Constant-time arrival and unlock paths. 

\Texticle{}Mitigates the \emph{multi-waiting} issue found in HemLock.

\Texticle{}Provides \emph{semi-private spinning}, similar to that found in TWA.  
A lock provides \emph{private spinning} if a most one thread busy waits on a given location
at a given time and the write to that location invalidates the cache of just that waiting thread.
Absent hash collisions in the waiting array, we have \emph{private spinning}, 
but, given the possibility of collisions, we instead have \emph{semi-private} spinning.  
As such, and because of the possibility of collisions, the worst-case Remote Memory Reference (RMR)  
complexity \cite{stoc08-attiya,podc02-anderson} is unbounded.  
If \emph{all} hapax values were to inadvertently hash to just one singleton slot, then operation 
devolves to the underlying Tidex algorithm, with \emph{global spinning}.  

\Texticle{}Hapax locks do not provide guaranteed local spinning. 



\Texticle{}Space efficient : Hapax Locks require one word of thread-local storage per thread to track the
current active private hapax block from which the thread is allocating; 2 words per lock instance for \texttt{Arrive} 
and \texttt{Depart}; one global word for the central hapax block allocator; and 4096 words -- 
shared over all locks and threads -- for the slots that form the waiting array.  
Any thread-local storage can immediately be reclaimed when the associated thread terminates.
The waiting array is immortal.  

\Texticle{}Directly amenable to modern polite address-based waiting on elements in the waiting array.
Examples of such techniques include \texttt{WFE} (ARM), \texttt{MONITOR-MWAIT} (Intel and AMD).

\Texticle{}Nested and imbalanced locking are naturally tolerated and handled gracefully.

\newpage 

\section{Discussion} 


\Invisible{Unstructured bag of bullet items; midden }  

\Texticle{}To preclude any potential concerns about roll-over in the hapax values, we expect implementations
to use 64-bit hapax values, making roll-over a practical impossibility.  

\Texticle{}To ensure constant-time doorway and unlock phases, 
Hapax Locks requires that the hardware supports constant-time 64-bit atomic loads, stores, exchange, and \texttt{fetch\_add}. 

\Texticle{}Like CLH, the algorithm requires only atomic loads, stores and exchange. 

\Texticle{}Unlike CLH, Hapax Locks can be trivially initialized with simple constants, avoiding the requirement of 
CLH \emph{dummy nodes}.  In fact Hapax Lock instances will normally be initialized to hold all $0$ values,
which can confer an advantage for initialization, and also allows global lock instances to reside in the 
in the \texttt{.BSS} or \texttt{.zero} program segment instead of the \texttt{.DATA} segment.  
Relatedly, Hapax Locks do not require explicit destructors.   
In addition, and unlike common CLH and MCS implementations, which cache queue nodes in thread-local storage,
no thread instance destructors are required.  (We note that MCSH\cite{MCSH} does not require such a thread-local cache
as queue nodes can be transiently allocated on-stack).  
  
\Texticle{}As there are no CLH- or MCS-like \emph{queue node} waiting elements, the algorithm
is not vulnerable to queue node migration, which can prove detrimental on NUMA systems, for CLH.

\Invisible{Coupling; dependencies; requirements; impose; impart ; } 

\Texticle{}TWA exhibits a tension on the waiting array size.  Larger arrays reduce hash collisions,
but, because of the undesirable ``marching around the array'' behavior, arising from the 
hash function and advancing ticket values, larger arrays could also increase cache footprint, 
cache pollution, and TLB misses.   The access pattern in the waiting array induced by TWA can also
trigger the processor's automatic hardware stride-based prefetcher, which itself can induce false
sharing and destructive interference. 
Relatedly, TWA provided no beneficial thread-level temporal re-use locality for accesses in the waiting array.  
In general, there is either no such trade-off, or a lessened trade-off with
Hapax Locks.  A bigger array is better, and reduces the odds of collisions or false sharing, although
the relative benefits of a larger aray tends to fade as we increase size.  

\Invisible{As noted above, TWA can exhibit an undesirable ``marching through the array'' behavior, which
is unfriendly to caches and TLBs.  Hapax Locks, depending on the design of the hash function, might
also ``march'' but as a much lower rate, providing some thread-level temporal locality 
in the array.}  

In Hapax Locks, however, locality in the array cuts both ways.  
For cache accesses, hash-induced locality acts to improve performance.  
But if we encounter collisions (or near proximal collisions that result in false sharing), 
then those collisions might persist for longer periods as a thread tends to re-access the same index until
it reprovisions its block.  

Assuming an ``informed'' hapax \texttt{ToSlot} hash function, a thread will exhibit strong temporal locality to a specific slot
in the array, which, in terms of simple cache hit rates, is desirable. 

\Invisible{Smart; informed; co-designed} 

\Texticle{}Mixing the lock address into the hash function will tend to mitigate the effect of \emph{multi-waiting}
as found in HemLock.  That is, Hapax Locks are less prone to multi-waiting than HemLock, although it can
still occur because of collisions. 

\Texticle{}Hapax values are globally and temporally unique -- 
one-time, one-shot, single-use, non-recurring, and disposable -- and map to a thread 
and a lock instance and a specific acquire-release episode by that thread on that lock. 
Non-recurring values, which are ephemeral and evanescent, allow the algorithm to avoid \emph{ABA}\cite{ArtOf} pathologies. 
Once a hapax value is installed into any \texttt{Arrive} field,
it is never reused in the lifetime of the process.  The values are strictly unique over that tenure.

\Invisible{Throw-away; one-shot; one-time; disposable; single-use; non-retrograde}  

\Texticle{}Once a hapax value is installed into any \texttt{Arrive} field,
it is never reused in the lifetime of the process.  The values are strictly unique
over that tenure.  For comparison, \emph{UUID}\cite{UUID,rfc9562} values, which are effectively 128-bit random numbers,
are unique only with very high probability. A related concept is \emph{Snowflake IDs}\cite{SnowFlakeID}, 
which incorporate a timestamp field and are also not strictly unique. 
We note that hapax values are effectively unordered, and only useful for equality comparison.    
(The current hapax allocation implementations happens to be ordered for a given intra-thread stream of values,
in the sense that a larger value must have been allocated after a smaller value.  Hapax Locks, however,
do not depend on that particular implementation property.)  

\Texticle{}Hapax values are sub-allocated from thread-local blocks of consecutive hapax values, and thread-private blocks,
when exhausted, are replenished by means of a global block allocator.  Larger blocks mean reduced
traffic on the global allocator. 

\Texticle{}In Ticket locks, for example, a thread waits for its \emph{own} ticket value to
appear in the lock's \texttt{Grant} (``now serving'') field, whereas in Tidex and Hapax Locks,
a thread waits for its predecessor's identity, obtained via the atomic exchange,
to appear in a variable associated with the lock. 

\Texticle{}The \texttt{Arrive} field contains the hapax identity of the most recently arrived thread
while \texttt{Depart} holds the hapax identity of the most recently departed thread --
the last thread to release the lock.                                   
The lock is in \emph{unlocked} state iff \texttt{Arrive} equals \texttt{Depart}. 
The identity of the current owner is implicit.      

\Texticle{}\red{\textbf{KILL:REDUNDANT}}As noted above, using hapaxes instead of 
fixed thread identities confers the following 
advantages.  First, in the classic Tidex algorithms, arriving threads need to have
at least 2 identity values, and must ensure that they don't install a value into
\texttt{Arrive} which happens to also reside in \texttt{Depart} as a residual value,
otherwise the lock inadvertently appear to not be held, resulting in exclusion failure.
To accommodate this requirement, Tidex, on arrival, fetches from the \texttt{Depart} 
field and switches to an alternative identifier, if the values matched.  
(Hapax Locks effectively provides an ``almost infinite'' stream of unique alternative identities).  
That fetch of \texttt{Grant} in Tidex adds to the path and coherence traffic generated 
by arriving threads.  
Second, threads waiting via the waiting array slots can recognize the specific identity
of their predecessor, and effect handover without needing to consult the \texttt{Depart} field.
In comparison, in TWA, which also uses proxy waiting with a waiting array, waiting
threads, when they observe the sequence number in the waiting array change, must
still ratify that they received ownership by rechecking the ticket lock \texttt{Grant} field
against their own ticket number.  This increases the handover latency and generates
additional coherence traffic, compared to Hapax Locks.  
Finally, Hapax Locks can update the waiting array with just a simple store of the 
hapax value, whereas TWA must use an atomic operation to update the waiting array
slot sequence numbers.  

\Invisible{conceptually: effective infinite unbounded stream of unique values = ticket fountain }  

\Texticle{}Compared to Tidex, Hapax Locks avoids the arrival redundant identity check,
allows semi-private waiting in the array, whereas Tidex requires global spinning.  
Hapax Locks offers a direct handover optmistic path that is more efficient under contention.  

\Texticle{}As hapax values never repeat or reappear, we can use them instead of 
simple TWA-style sequence numbers -- as change indicators -- in the waiting array to indicate that the
waiter may need to recheck the \texttt{Depart} field to confirm if 
ownership was conveyed to it. 

\Texticle{}For \Hapax{} with invisible waiters, in the case of hash collisions in the waiting array, a hapax value can
be overwritten by a colliding access.  In this case the thread waiting for
one value might miss the direct handover optimization, but it will still observe
the change and recheck the \texttt{Depart} value, recognizing handover in that fashion.
Crucially, once a hapax value appears in the array, it will never reappear anywhere or at any time 
in the array.  That is, hapax identity values are \emph{non-recurring} and slot values thus never \emph{flicker}
back to a previous state. 

\Texticle{}The odds of collision in the array is equivalent to 
the so-called \emph{Birthday Paradox}\cite{BirthdayParadox}.  
\Invisible{Equivalent to; Governed by Birthday Paradox} 

\Invisible{Collision taxonomy : Inter-thread intra-lock collisions vs inter-lock collisions}  

\Texticle{}We observe collisions in the wait array can be either \emph{intra-lock} or \emph{inter-thread}. 

\Texticle{}Hapax Locks with invisible waiters can allow threads in different processes to 
safely synchronize on shared inter-process locks, as long as a shared singleton nonce allocator is used, and 
the waiting array is mapped into each address space.  This is enabled by virtue of the fact that 
no addresses are shared between participating threads.   

\Texticle{}Hapax Locks are not \emph{context-free} as we need to communicate an acquire-release
episode's hapax value -- installed by the atomic exchange in the \texttt{lock} method --
to the corresponding \texttt{unlock} method.  We note that CLH, MCS, and Reciprocating locks all require context,
while HemLock, Ticket Locks, and TWA are context-free.  

In managed runtime environments such as the HotSpot Java Virtual Machine, with balanced locking 
under the ``synchronized'' construct, such context information can be kept in the stack frame.  In C++,
context can be kept in \texttt{std::lock\_guard} or \texttt{std::scoped\_lock} instance, 
which is typically allocated on stack.  Rust's \texttt{Mutex<T>} is similar.  
Such constructions are an example of the the \emph{Resource Acquisition is Initialization(RAII)}\cite{RAII} idiom.
Context can also be conveyed to the unlock operation via a variant API, which allows passing of information,
by using a lambda-based interface, through extra fields in the lock body, or via data structures held 
in thread-local storage that track held locks.  

\Texticle{}We would like to avoid collisions in the waiting array, but, for the purposes of performance,
would also prefer to avoid near- or \emph{proximal}-collisions in order to reduce spatial false sharing.   

\Texticle{}In sensitivity experiments with Hapax Locks with the number of ready threads set to more than the number of CPUs 
(intentionally forcing the system into preemption) and with the waiting array intentionally set to be small,
we find significant rates of collision and sharing of slots in the waiting array.  
Many of the waiters are preempted, however, and not actively busy waiting.  
While collisions degrades performance because of the failure to provide private spinning, performance 
is actually dominated by the fact that we need kernel preemption to make forward progress, and preemption 
operates in geologic time.   Assuming FIFO succession (or really, any form of succession that employs direct handoff), 
the choice of lock algorithm doesn’t play much of a role in performance in this regime of operation, 
when kernel preemption is in play.   
Performance abruptly falls off at the onset of preemption.  
There is not, for example, any appreciable performance difference between ticket locks (global spinning)
and CLH (private spinning) at that point.   
The only locks that can withstand that kind of situation, without completely melting down, are those that use some type of
bounded barging bypass or concurrency restriction, such as GCR, Fissile, etc.   
This tells us that we really only need to size the waiting array as a function of the number of the CPUs, 
and not the number of expected participating threads.     

\Texticle{}One downside to Hapax Locks is, like TWA, we can not guarantee that the locations being
busy-waited upon in the array are homed on the same NUMA node as the waiting thread.  In contrast,
MCS and Reciprocating Locks provide \emph{locally homed spinning}, where the spinning thread and the memory location
being waited upon reside on the same NUMA node, which can yield a performance benefit on platforms
such as Intel NUMA systems with the UPI\cite{UPI,icpp23-katevenis} fabric which uses \emph{home-based coherence}.  
In home-based coherence, all coherence probes on a given cache line are sent to the line's home node, 
which adjudicates and arbitrates access. 
This approach can increase latency as more ``hops'' might be required, if the home needs to interrogate other nodes,
but can reduce bandwith by reducing the need for broadcasts.  
On other common platforms, including Intel QPI, AMD, and ARM, performance is less sensitive to the location of the home node 
of a waited-upon location relative to the placement of the spinning thread.  Instead, performance on those platforms,
-- coherence miss costs -- is a function of the location of the requester, in the topology, the type 
of request (read or write), and the location of other caches that have the line, and the status of the
line in those caches. 

Similarly, and for the same underlying reason, TWA locks\cite{EuroPar19-TWA,arxiv-TWA} perform well on
all platforms except Intel NUMA UPI, where they lag behind other locks that use private locally homed spinning. 

In Hapax Locks, the slot in the array where a thread waits for the predecessor's value to be announced,
signaling that the predecessor abdicated,  is a function of the predecessor's hapax value.
As such, absent extra indirection, we do not have control where a thread waits -- where the location in the array is homed.  
This is similar to CLH and HemLock, but unlike MCS and Reciprocating Locks, where waiting will be performed
on locally homed locations.  

\Invisible{Remote spinning = NOT locally homed spinning; potentially remote;} 

\Invisible{Where else there might be caches which have the line and require invalidation or transfer} 
\Invisible{MESI; MESIF; MOESI} 

\Texticle{}Hapax Locks (with invisible waiters)  provides \emph{dual} handover encoding for succession on contended locks with waiting
threads.  The preferred mode is via expedited transfer via the waiting array while the fallback is via
the \texttt{Depart} field. 

In more detail, Hapax Locks effect contended lock ownership handover via 2 distinct modes in the \texttt{unlock} path. 
Recall that the outgoing thread, executing in \texttt{unlock}, updates the \texttt{Depart} field
and then updates the associated waiting array slot with its own hapax values.  
If the waiting thread observes the waited-upon value appear in the slot in the array, it
can exit the waiting phase immediately and enter the critical section, without
any need to confirm the value of \texttt{Depart}.  But if the slot in the array
shifts to any other value, say $X$, the waiter needs to recheck \texttt{Depart}.
If \texttt{Depart} matches the waited-upon value, the thread has been granted ownership and 
can enter the critical section.  Otherwise, the waiting thread will resuming waiting
on the slot value and continue to wait while the value remains equal to $X$.
When the value next shifts,  thread will again check for expedited handover, and otherwise 
recheck \texttt{Depart}, and, if necessary reenter the waiting loop using the recently observed value. 

The waiting thread needs to triage 3 potential cases, while polling (busy-waiting) on the slot.
If the value did not change since it was last fetched, the thread just continues to spin.
If the value changed to waited-upon value, the waiting thread is now the owner, via
direct expedited handover, and can enter the critical section.
Othewise, the value in the slot shifted, so the waiting thread must conservatively 
recheck \texttt{Depart} before resuming the waiting protocol on the slot. 

\Texticle{}The particular associated slot in the array is determined via a hash on the hapax value, implemented in 
the \texttt{ToSlot} function. 

\Texticle{}Under contention, the value of the current outgoing owner's \texttt{hapax} variable, 
executing in \texttt{unlock}, is the same as the incoming successor's \texttt{pred} variable, as found 
in the successor's \texttt{lock} phase.  

\Texticle{}Hapax Locks are \emph{thread-oblivious}\cite{topc15-dice} in that one thread can acquire
a lock, but a different thread, in possession of the associated context, communicated from the 
acquiring thread,  can subsequently release the lock.  As such, they can be used in
NUMA-aware Cohort Locks, as bounded binary semaphores, or other use cases where thread-oblivious locks are necessary. 
In addition, with the uptake of asynchronous futures, it is more common to find situations
where one thread accquires a lock, and some other thread may subsequently release that lock
\footnote{\url{for example : https://github.com/serenedb/serenedb/blob/main/libs/basics/unshackled\_mutex.h}}.  

\Texticle{}In the code in the listings, we use 128-byte alignment and padding to 
reduce false sharing.  The unit of cache coherence on Intel x86 is 64 bytes, but because
of the hardware adjacent sector prefetch mechanism, we conservatively elect to use 128 bytes.  
We intentionally did not sequester or isolate \texttt{Arrive} and \texttt{Depart} as they
tend to be accessed contemporaneously by arriving threads,  exhibiting spatial locality.   
As such, the lock body a Hapax Lock is extremely compact.  

\Texticle{}Implementing \texttt{try\_lock} is difficult if not impossible for Tidex.
But under Hapax Locks, given non-recurring values, and the consequent elimination of \emph{ABA}\cite{ArtOf} pathologies,
a thread in \texttt{try\_lock} could inspect \texttt{Arrive}, and then \texttt{Depart}, and if the values were
found equal, know with certainty that the lock is free, and the proceed to use an atomic compare-and-swap
to try to install a new hapax value into \texttt{Arrive}, over the previous value.  If the compare-and-swap
was successful, the arriving thread has acquired the lock, otherwise, some other thread arrived
and acquired the lock, and the \texttt{try\_lock} operator can return the expected failure indication.  

Relatedly, \texttt{try\_lock} is difficult to implement in CLH and in ticket lock forms where the
\texttt{ticket} field is only 32-bits and vulnerable to roll-over and inadvertent false-equal aliasing.  

Consider implementing \texttt{trylock} in Tidex.  We fetch and inspect the \texttt{Arrive} and 
\texttt{Depart} variables and find them equal, so the lock is \emph{apparently} unlocked -- 
\emph{apparently} as we didn’t get an atomic snapshot of the two variables, as we used two loads.  
Ignoring the alternative ID issue, we then use CAS to try to swing \texttt{Arrive} to our 
thread’s identity.  But the fields could have shifted underneath us,  from other concurrent actions 
on the lock, yielding an ABA issue where our CAS was successful, but we then discover that \texttt{Depart} 
is changed, in which case we’d be forced to wait in \texttt{trylock}.  We might also try to revert and
back-out our update of \texttt{Arrive}, with another CAS, but there is no guarantee of success as
other arrivals may have updated \texttt{Arrive}.  
The same scenario can manifest in ticket locks, although it takes 4G roll-over of 32-bit 
\texttt{Ticket} and \texttt{Grant} fields to occur.   If we are willing to pack and collocate 
\texttt{Arrive} and \texttt{Depart} fields (or \texttt{Ticket} and \texttt{Grant}) fields 
together, and then use a double-wide CAS, you can safely implement \texttt{trylock} in that fashion.  
If we adapt ticket locks to use 64-bit fields, then, as roll-over and ABA are impossible, 
as a practical concern, we can safely implement \texttt{trylock}, where the operator
first fetches the \texttt{Grant} field, observing value $G$, and then tries to optimistically 
apply atomic compare-and-swap to swing the \texttt{Ticket} field from $G$ to $G+1$.  

\Hapax{} also gives us a viable \texttt{trylock} as we have 64-bits and non-recurring values.  
We further observe that \texttt{trylock} is safe even with the \Hapax{} visible waiters and 
positive handover.  During positive handover episodes (sustained contention), we skip 
updating the \texttt{Depart} field (we resume updating \texttt{Depart} when contention subsequently abates) 
but we also know that \texttt{Arrive} and \texttt{Depart} 
must be different during such an episode, so the lock appears in \emph{locked} state, 
and any \texttt{trylock} attempt will fail.     

\Texticle{}Like TWA -- which is a variation of ticket locks augmented with a waiting array, to
avoid global spinning --  Hapax Locks are \emph{value-based} and avoid passing shared addresses between threads.  
TWA and Hapax Locks also share similar RMR complexity by virtue of the shared waiting array.  
Unlike TWA, Hapax Locks provides expedited handover under contention.  

\Texticle{}In TWA and Ticket locks, a thread knows that its immediate waiting successor must hold ticket value
of just one more than the ticket value allocated to the current owner.  In Tidex and Hapax Locks, however,
the departing thread has no particular knowledge about the successor, other than that it exists.  

\Texticle{}We could also employ \texttt{futex} address-based waiting, in the waiting array,
but note that hapax values in the array
must occupy 64 bits whereas the Linux kernel futex interface permits waiting only on 32-bit locations.  
Simply waiting on the low-order truncated 32-bits of a hapax is unsafe and could result in missed 
wakeups and progress failure.  
Instead, we could use the \emph{waiting chain} construct in \cite{TWASemaphore} where
each slot in the array would be provisioned with an additional pointer to the head
of a chain of waiting elements, implemented as \emph{concurrent pop-stack}.  
Threads waiting on those chains would wait (using a futex) on a simple private flag variable
in the elements they pushed onto the stack.  

In the case of futex-based waiting, as we have semi-private waiting 
in the array, if a Hapax Lock is highly contended, then the set of futex addresses waited upon
are distributed over the array, which, in turn, acts to diffuse kernel contention over the set of 
futex sleep channels, further improving scalability, as those sleep channels themselves are 
protected by kernel spin locks.  

\Texticle{}Hapax Locks are trivially amenable to using more sophisticated waiting techniques 
on the slots, such as \texttt{WFE} on ARMv8, \texttt{MONITOR-MWAIT} on Intel, 
or \texttt{futex}-based waiting.  On ARMv8 systems we have observed that waiting via 
\texttt{WFE} makes a signficant improvement in contented throughput, compared to \texttt{YIELD},
for locks that uses private spinning, but that \texttt{UMONITOR} and \texttt{MONITORX} made
no appreciable improvement for throughput on AMD and Intel processors.  

\Texticle{}We observe in passing that hapax values could be retrofit into the TWA locking algorithm.  
In \texttt{unlock}, instead of using an atomic \texttt{fetch\_add(1)} in the waiting array slot,
we could instead conjure a hapax value and simply store that value into the waiting array slot. 

\Texticle{}Hapax values could also be basis for \emph{counter-based random number generators}\cite{CBRN}.

\Invisible{There are a number of such issues where Hapax Locks look bad in theory, but in practice they are just fine, 
making for a systems vs. theory dichotomy and tension.} 
\Invisible{Best effort; as-if; trylock allowed to return failure if lock was held at any point during its activation}  

\Invisible{Carrier thread; mount; dismount} 
\Invisible{Register; emplace; install; ensconce; announce; publish-subscribe;} 
\Invisible{Entangled; Leak; Escape; exposed; publish;} 

\Invisible{tension; trade-off; bet; compromise; strike a balance}  
\Invisible{ABA; flicker; transient; overwrite; stomp; squash; quash; } 
\Invisible{Arrive=Ingress; Depart=Egress}

\Invisible{Rosetta stone -- Mapping from public paper lock names to source code algorithm C++ class names : 
see migro-MutexBench.cc NonceLock HapaxLock and Exotic03NT family} 

\Texticle{}Many platforms exhibit \emph{topological favoritism} or, specifically, \emph{NUMA favoritism},
where atomic operations that are either \emph{near} (in the topology) the home node of the
cache line or are near the cache that currently holds the line in modified state are
more prone than \emph{far} requesters to be successful and serviced in the near time.
Perversely, test-and-set locks are \emph{NUMA friendly} on such platforms in that they reduce
\emph{lock migration}\cite{topc15-dice,PPoPP12-dice,arxiv-CNA,EuroSys19-CNA} by
virtue of such architectural unfairness, offering improved performance compared
to pure FIFO locks.  Backoff mechanisms commonly used in test-and-set locks
also exacerbate unfairness, as a waiting /thread that has backed off is less like to acquire
the lock in unit time than a more recently arrived lock.  Backoff also represents
``dead time'' and is not work-conserving, as it is possible, under varying load,
that the last waiting thread is stalled in a long backoff period when it could otherwise
acquire the lock and drive progress.  

\Texticle{}We believe it is safe to aggressively fetch \texttt{Depart} early in the \texttt{lock} operation,
before performing the atomic exchange on \texttt{Arrive}.  That is, we can lift or hoist
the fetch of \texttt{Depart}, which normally apears after the exchange, to before and ``above'' the exchange.  
Such an optimization, with 
early fetch of the \texttt{Grant} field, is also safe in Ticket Locks if we use 64-bit
\texttt{Ticket} and \texttt{Grant} fields.  

\Texticle{}We call out a confounding benchmark phenomenon we call the \emph{re-arrival} effect.  
We say a lock algorithm is \emph{unlock-terse} if, under sustained contention, it avoids updating 
lock body fields in \texttt{unlock}, and \emph{unlock-prolix} if it usually updates the lock 
body in \texttt{unlock}.  Unlock-prolix lock algorithms that always write into the shared lock body, 
in \texttt{unlock}, such was Ticket Locks, will usually suffer more coherence traffic 
than do unlock-terse algorithms -- such as MCS and CLH -- that avoid, under sustained contention, 
updating the global lock body state in \texttt{unlock}.

Say we have a maximum contention benchmark with an empty non-critical sections, and sustained contention.  
Threads release the lock in \texttt{unlock} and then immediately re-arrive in \texttt{lock} to re-acquire the same lock.
But, with an empty critical section, and assuming all the lock body fields are collocated on the 
same cache line, the updates by an unlock-prolix algorithm in \texttt{unlock} will drive the 
line(s) underlying the lock body into M-state, and the subsequent \texttt{lock} 
operation will re-access those same lines while they still reside in the local cache in M-state, 
as no other lock operations have had an opportunity to arrive or depart and invalidate the line.  
This effect somewhat distorts performance and confers a relative performance benefit to prolix lock algorithms
in the case of maxmimum sustained contention with no intervening non-critical section.  
The degree of impact is specific to the architecture and the lock algoirthm -- whether prolix or not.  

The \texttt{lock} operator, for an unlock-prolix algorithm, can avoid coherence misses that it might otherwise incur
if threads arrived between the \texttt{lock} and \texttt{unlock}.  

We note that the existence of lock context passed through the lock body can trigger the re-arrival effect,
even if the algorithm is normally unlock-terse. 

Using randomized non-critical section durations acts to mitigate this effect, and is also
arguably more faithful to the real-world application behavior we are trying to model in benchmarks. 

\Texticle{}Similar to CLH and Reciprocating locks, no explicit queue of waiting threads is made 
manifest.  Waiting threads know only the identity of their immediate successors.  As such, these
locks can not trivialy be adapted, when implemented within a pthread mutex-condvar environment, to 
enable the classic \emph{wait morphing} optimization, where signaling a condition variable
within a critical section simply moves threads from the condition variable's \emph{wait set} to
the mutex's queue of waiting threads.  

\Invisible{distorts, pitfall, confounding factor} 

\Invisible{Terse vs talkative} 

\Invisible{back-to-back, no intervening} 

\Invisible{Frugal, parsimonious, profligate, write-heavy, prolific, profligate, 
terse, Hypergraphic, logorrhea, scriptive, terse, grandliquotent, prolix, talkative} 

\Texticle{}Another benchmarking pitfall is as follows.  
Say we have a FIFO lock algorithm, and, for the purposes of this explanation, an empty critical section.
It it then common to find a long-standing repeating cyclic admission schedule.  Often, this 
schedule is set via happenstance at startup, but then persists for long periods. 
A given schedule can be either favorable or unfavorable, for throughput, in terms of the system
topology.  Using NUMA as an example, a given schedule might reflect either many or few \emph{lock migrations},
which in turn will reflect in the throughput achieved by that schedule.  
But other topology structures, such as AMD's CCX clusters, can yield similar favoratism.  
This, in turn, can result in higher run-to-run variance, if we assume the initial schedule
arrises in a mostly random fashion.     
This effect can be mitigated in part by injecting randomized non-critical section delays, in order to
shuffle or shift the schedule and avoid long-standing persistent schedules.

\Texticle{}TWA vs Hapax.  
Naively, we might expect the baseline \Hapax{} (with invisible waiters -- the baseline algorithm) to 
yield about the same performance as TWA.   But Hapax often underperform TWA for the following reasons: 

For the hapax forms, we need to pass the hapax value we installed, as \emph{context} from 
a \texttt{lock} operation to the corresponding unlock() operation.  In the \texttt{LD\_PRELOAD} 
implementations, we keep any lock context as an extra field in the lock body, but unfortunately 
accessing the context field generates coherence traffic.  But for TWA (like ticket locks), we don’t 
need to pass context, but can instead just increment (non-atomically) the \texttt{Grant} 
field in the lock body.

TWA enables an optimization where we let the thread that’s the immediate successor 
shift from waiting in the waiting array to direct spinning on the \texttt{Grant} field in the lock 
body, hopefully yielding a more expeditious handover.   
Whether this specific optimization is profitable or not is very platform-specific and related to coherence costs. 
The ticket value a thread has in-hand gives a sense of proximity to the front of the queue.  
There’s no such optimization available for the Hapax family.   The hapax value doesn’t really 
convey any sense of proximity to the ``front'' position in the logical queue of waiting threads, 
whereas a ticket value does.

TWA, however, exhibits the ``marching around the waiting array'' effect, which is unfriendly to caches.

\Invisible{Arrival; Arrival; Ingress; LastToArrive; MRAT; Ingress; Incoming} 
\Invisible{Egress; Grant; Outgoing; Depart; Departure;  Drain; Exeunt; Relinquish; Renounce; } 

\Invisible{Ratify; Validate; confirm; recheck; } 
\Invisible{keywords: region; sector; block; sequence; zone; 
zipcode; area code; country code; geocode; 
Ticket; Token; Identifier;    
Dispenser; factor; stream; generator; 
Nonce Hapax; private unique identifier stream; ticket stream; } 
\Invisible{sub-allocate} 
\Invisible{Profluent} 
\newpage 

\section{Optional Optimizations} 

\Invisible{Optional optimistic optimizations opportunities ; embellishments; variations }  

\Texticle{}We can intuit thread-vs-thread collisions when a waiting thread observes
the value in the waiting array shift, but not to the waited-upon value.  In this case,
to reduce the odds of further collisions at that same waiting array index, the thread
could optionally abandon its current block of hapax values, which are known to collide, and, the next time it allocates a hapax,
allocate an entirely new block.  While this technique for collision reduction is viable, 
it consumes hapax values at a higher rate and might eventually result in premature hapax roll-over.  
As an option, a thread might hold 2 active blocks and switch between them based on collisions.


\Texticle{}There are some other approaches we could use to reduce waiting array collisions.  
One thought would be to just use an arbitrary ``proper'' hash that’s not hapax-aware.   
Intra-thread temporal access locality in the array is much reduced.   Another idea would be to hash on the lock address, the thread ID 
part of the hapax, and the sub-sequence part of the hapax, but throwing away, say, the lower 5 bits 
of the local sequence number.   So a thread would go to the same slot in the array 32 times in a row, 
which should still confer some locality benefits, but any specific thread-vs-thread collision 
scenarios would persist for only 32 episodes.   This constitutes a reasonable compromise between
collision penalties and cache locality benefit. 

\Texticle{}To further allay any residual concerns about hapax overflow we might shift to a hapax formulation
that has a 64-bit thread ID field and a 64-bit sub-block value.   At that point all concerns about overflow are eliminated.   
Modern Intel x86 and AMD processors now guarantee that properly aligned 128 byte loads and stores are 
completely atomic via the \texttt{movdqa} instruction.  Using 128-bit hapax values would increase 
the lock instance size.  Also, the Intel instruction set only has \texttt{lock:cmpxchg16b} for 128 byte atomics.  
There is no 128-byte atomic exchange.  So we’d need to emulate the exchange with a CAS loop, and we lose our 
constant-time arrival \emph{doorway} step and are potentially more vulnerable to architectural induced unfairness.   
Modern ARM also has 128-byte loads and stores and LL-SC.    
Encoding tuples in this fashion provides a safe solution that is immune to roll-over but which is not viable 
commonly available current processors.

\Texticle{}The following variant of the Hapax algorithm works well in a managed runtime environment with garbage collection, 
like that provided by the HotSpot Java Virtual Machine.  
Instead of conjuring up our hapax values, we just use a reference to a freshly instantiated object as the hapax identity :  
\texttt{hapax = new Object()}.   These objects will typically be allocated out of the thread-local allocation buffers 
(TLABs) via a bump pointer, so in the normal case such allocation is extremely fast and requires no synchronization.  
The astute reader may notice the analogy between hapax blocks and TLABs.  
Garbage collection also ensures the hapax values are temporally unique, even in a 32-bit JVM. 
Of course this technique offloads some of the problem off onto the garbage collector and object allocator.  
Allocating hapax values consumes space, which must eventually be recovered.  
The arrival path is no longer constant-time.   In addition to possibly needing locks, it could 
in theory even induce a full garbage collection.  Overflow is impossible, though.  

\Invisible{At this point we might as well busy-wait on a field within the predecessor's node, resulting in approach
that's closer to CLH.} 

\Texticle{}Depending on the hapax allocation scheme and hash function, which maps hapax values
to waiting array indices, if a given block, or portion of a block, maps to just one slot, then we can precompute and ``memoize'' 
the slot index and store the value in thread-local storage and compute the hash just once per block.

That is, depending on the block size, and specifics of the hash function, in \texttt{ToSlot}, 
which maps hapax values to indices (slots) in the waiting array,  a thread may be able to precompute
and cache in thread-local storage -- or ``memoize'' -- the address of the waiting array slot it will use for a sequence of
hapax values, avoiding some computation. 

\Texticle{}Given that uncontended locking is extremely common, to reduce the consumption and \emph{burn date}
of hapax values, we might institute optimizations to recover values from uncontended locks, where those
values never leaked or were exposed to other threads.  A thread could maintain a simple cache of hapax values
it recovered from re-locking the same lock instance.  While workable, we do not currently believe
the bookkeeping overheads associated with this specific optimization make the technique profitable.  

\Invisible{Recapture; reclaim; reuse; recover} 

\Texticle{}If we expect contention to be rare, then we can apply an optional optimization in the 
\texttt{unlock} path.  After storing in the \texttt{Depart} field, a thread will then fetch from \texttt{Arrive}.
If the values are the same then we have reverted to \emph{unlocked} state and can safely elide the store
into the waiting array.  This optimization reflects something of a bet or trade-off.  If we try to apply
the optimization and fetch from \texttt{Arrive} but find contention (the fetched value differs from the 
value we just stored into \texttt{Depart}) then we still need to store into the waiting array, but have
generated additional futile coherence traffic by fetching from \texttt{Arrive}, to no avail.

\Invisible{This optimization reflects; constitutes; embodies...} 

A reasonable heuristic is a follows.  If, when we acquired the lock, our thread had to wait, then we
can assume prior contention on a lock accurately predicts future near-term contention.  
In that case, we bet that contention still exists, and update the waiting array unconditionally.
But if we acquired without waiting then, in the corresponding unlock operation,  we try to apply the 
optimization and condition the store into the array based on the value fetched from the \texttt{Arrive} field. 

\Texticle{}Listing-\ref{Listing:Nonce-Allocator} describes a variety of optimizations that can be applied to
the hapax block allocation procedures.  Some of the techniques therein might be useful in environments where
thread-local storage is not available or inefficient, and hapax values would need to be allocated one-at-a-time.

\Texticle{}The \emph{visible waiters} \Hapax{} form can optionally optimize uncontended \texttt{unlock} 
operations as follows.   If, in \texttt{unlock}, we find that \texttt{Arrive} still equals the hapax
value installed by our thread, in \texttt{lock}, then there are no threads waiting on the lock, and we have 
a simple uncontended \texttt{unlock} operation, so we can safely use CAS 
to try to attempt to swing \texttt{Arrive} back to the value found in \texttt{Depart},
If the CAS is successful we have rolled back the state to \emph{unlocked} and can then return without any 
need to update the waiting array slot.  
That is, instead of updating \texttt{Depart}, we roll-back and revert \texttt{Arrive}.  
Furthermore, if desired, as that hapax value was thus never observed by other threads, we can
reuse it for subsequent operations.  
\Invisible{Revert; roll-back} 

\Invisible{Smart; informed; co-designed; intentional design; }

\newpage 

\section{Optimized Hapax Allocator Variations} 

\lstset{caption={Hapax Allocator Variation}}
\lstset{label={Listing:Nonce-Allocator}}
\begin{lstlisting}[mathescape=true,escapechar=\%]
// Allocate a new never-before-seen Hapax identity value 
//
// We have a variety of options available for picking a
// lane from which to allocate : 
// Variations:
// @  Pick lane random uniform in [0,NLanes)
//    This diffuses coherence activity over the array and avoids
//    situations where the allocator itself might constitute a hot-spot.
//    In turn, this make smaller blocks more practical by reducing
//    the cost imposed when a thread needs to reprovision.
// @  Use the CPUID as as the lane and provision the array with one lane per CPU.
//    Index into the array using the CPUID, where all array
//    elements are sequestered as the sole member of a cache line.
//    Assuming sequestered fields, this virtually eliminates false
//    sharing and coherence traffic, as a field will be accessed
//    only by one CPU.
//    We could also consider arranging the array of generators
//    such that the elements are then co-homed with the NUMA nodes that 
//    will access those elements.  
// @  Pick lane as a function of CPUID -- hash on CPUID
// @  Pick lane as a function of NUMA node ID of current thread
// @  Pick lane as a function of the current time, assuming
//    the clock source has sufficient resolution as to avoid collisions.  
// @  Pick lane as a function of thread identity or, equivalently, perhaps 
//    a hash on the address of a TLS value on on-stack variable
// @  Use hybrid combination of the previous
// @  Pick lane as a hash function of the previous nonce value used by 
//    this thread, or, equivalently, the previous BlockBase used by this thread.
//    This is tantamount to implementing a thread-local Counter-based PRNG (CBPRNG). 
//    https://en.wikipedia.org/wiki/Counter-based_random_number_generator
// @  We might also partition the set of lanes geographically, by NUMA node ID,
//    resulting in a topology-aware allocator.  
//    NUMA nodes would be assigned to various contiguous sectors in the 
//    Lanes[] array.  We could then pick a lane within that NUMA sector
//    by any of the polices mentioned above.  
//    This also acts to encode the NUMA node ID into the hapax field,
//    forming a geographic "zip code" embedded in the identifier
//    Such the hapax values convey geographic position in the system
//    topology via the CPUID or NUMA node ID of the associated thread.
//
// The techniques above are useful, and reduce the cost a threads incurs
// to reprovision, allowing smaller block sizes to be more practical.
// But if the block size is sufficiently large to amortize the cost of
// allocating new blocks from a singleton global allocator, then little benefit
// will be gained from such tactics, as the global allocator 
// will be accessed infrequently.
 %\IRule%  
static uint64_t NextHapax () { 
 %\IRule% // To reduce false sharing, we use alignas() to sequester Base
 %\IRule% // as the sole field on its underlying cache line. 
 %\IRule% // Or, if we have a sufficient number of lanes, and NLANES is large, 
 %\IRule% // we could skip sequestration and isolation, and just let normal diffusion
 %\IRule% // of access probabilities over the set lanes act to mitigate false sharing
 %\IRule% struct Lane { 
 %\IRule%  %\IRule% std::atomic <uint64_t> Base alignas(128) {0} ; 
 %\IRule% } ;
 %\IRule%  
 %\IRule% // Both NLANES and BlockSize are tunable parameters 
 %\IRule% // Larger BlockSize values act to reduce how often a thread
 %\IRule% // needs to access the global allocator by letting a thread
 %\IRule% // sub-allocate privately, from a block it keeps locally in hand. 
 %\IRule%  
 %\IRule% // We expect the ToSlot() operator to be intimate with hapax
 %\IRule% // allocator, so it can map identifiers to waiting array slots in a way
 %\IRule% // that emphasizes intra-thread cache locality and reuse but also acts
 %\IRule% // to reduce collisions and false sharing between threads. 
 %\IRule%  
 %\IRule% static constexpr uint32_t NLANES = 4 ;                
 %\IRule% static_assert (NLANES > 0 && (NLANES & (NLANES-1)) == 0) ; 
 %\IRule% static Lane Lanes alignas(128) [NLANES] ; 
 %\IRule% static constexpr uint64_t BlockSize = 64*1024 ;      
 %\IRule% static_assert (BlockSize > 4 && (BlockSize & (BlockSize-1)) == 0) ; 
 %\IRule% // We desire lock-free operation for performance and progress properties
 %\IRule% // but, not for correctness
 %\IRule% static_assert (std::atomic<uint64_t>::is_always_lock_free) ; 
 %\IRule% static constinit thread_local uint64_t PrivateHapaxBase = 0 ;     
 %\IRule%  
 %\IRule% auto hapax = PrivateHapaxBase ++ ; 
 %\IRule% if ((hapax & (BlockSize-1)) == 0) [[unlikely]] { 
 %\IRule%  %\IRule% // crossed edge of block allocation 
 %\IRule%  %\IRule% // current block is exhausted so must reprovision
 %\IRule%  %\IRule% // Allocate a block of unique hapax identity values 
 %\IRule%  %\IRule% // This check and path also handles per-thread initialization
 %\IRule%  %\IRule% auto lane = RandomInteger() & (NLANES-1) ; 
 %\IRule%  %\IRule% auto u = Lanes[lane].Base.fetch_add(1) ; 
 %\IRule%  %\IRule% auto BlockBase = BlockSize * ((u * NLANES) + lane) ; 
 %\IRule%  %\IRule% // We should check for integer overflow!  
 %\IRule%  %\IRule% // By convention, we never allocate a hapax having value 0
 %\IRule%  %\IRule% assert (BlockBase > hapax) ; 
 %\IRule%  %\IRule% hapax = BlockBase + 1 ; 
 %\IRule%  %\IRule% PrivateHapaxBase = BlockBase + 2 ;      // resume allocation stream at +2
 %\IRule% }
 %\IRule% assert (hapax %$\neq$% 0) ; 
 %\IRule% return hapax ;
} 


\end{lstlisting}  

We could, as an optional optimization, provide a small array of global ID generators, 
initialized with properly skewed values to ensure uniqueness, and when we need to reprovision, pick one ``lane'' at
random and allocate from that element.   This diffuses coherence activity over the array and
avoids the situation where the allocator might itself constitute a hot-spot. 
In turn, this would make smaller block sizes more practical, reducing the cost a thread
needs to reprovision.  
There doesn’t seem to be any practical benefit for this optimization, however, if the block size is sufficiently 
large to amortize the cost of allocating new blocks from a singleton allocator, as it will be accessed infrequently.
Listing-\ref{Listing:Nonce-Allocator} shows an implementation of such a hapax allocator.

Instead of randomly picked slots, another viable method is to arrange the array of generators to be 
indexed with the CPUID or NUMA node ID of the allocating thread.  This greatly reduces coherence traffic, assuming
each generator is properly sequestered on its own cache line or sector.  Interestingly,
such hapax values also convey \emph{geographic} position in the system topology -- via the CPUID or NUMA node ID --
of the associated thread, 

\newpage

\section{Hapax Lock Variation with Visible Waiters} 

\lstset{caption={Hapax Lock Variation with Visible Waiters}}
\lstset{label={Listing:Nonce-Visible-Waiters-FULL}}
\begin{lstlisting}[mathescape=true,escapechar=\%]
// HapaxLock with visible waiters
//
// @  Visible waiters in waiting array slots
// @  Dual representation : 
//    Depart is authoritative-definitive ground truth 
//    We employ proxy waiting in wait array
// @  Each visible waiter occupies a slot in the waiting array
// @  Amenable to futex address-based waiting
// @  Falls back to either classic Tidex or HapaxLock in the event of hash collisions in the waiting array
// @  Normally slots in the array are 0
//    Threads arriving to wait try to CAS the slot associated with the waited-upon value (which is non-0)
//    from 0 to that value.
//    If that CAS fails, because the slot is occupied by another waiter,
//    they revert to degenerate Tidex global spinning.
//    Otherwise that waiting thread owns the slot for the duration of the their waiting phase
//    and busy waits for the value to change to _any other value.
//    The corresponding unlock() operator tries to clear the slot back to 0 with CAS.
//    The slot value can "flicker" to 0 and then to other values, because of concurrent activities,
//    so the reader (waiter) is safe to interpret any change in the value as the matching unlock().
//    This is safe by virtue of the hapax value non-recurring property.
//    The slot can not ever revert to a previous value, so we do not miss wakeups to a waiter.  
// 
// @  Benefits
//    *  Faster expedited fast-path direct handover
//       The waiting successor does _not need to fetch the global Depart field to confirm handover.
//       Shorter critical path during lock passage 
//       Reduces coherence traffic on Depart 
//    *  Reduced write coherence traffic on Depart field
//       We can elide updates to the global Depart in unlock() if we can accomplish direct handover
//       If unlock() was able to confirm handover to the correct waiter then can safely skip updating Depart.
//       Requires postive handoff handshake via visible waiters
//    *  Overall, the contended handover critical path is very tight for both
//       the outgoing thread in unlock() and the waiting successor. 
// 
// CONSIDER:
// @  Fallback to crude global waiting Tidex or to HapaxLock with invisible waiters ?!
//    Currently we just revert to Tidex, but we could do better.  

struct HapaxLockVisibleWaiters { 
 %\IRule% using Hapax = uint64_t ; 
 %\IRule%  
 %\IRule% struct Slot { 
 %\IRule%  %\IRule% Atomic <uint64_t> VisibleWaiter {0} ; 
 %\IRule%  %\IRule% Atomic <uint64_t> _useq {0} ;            
 %\IRule% } ; 
 %\IRule%  
 %\IRule% // We require lock_free for performance, not correctness
 %\IRule% static_assert (std::atomic<uint64_t>::is_always_lock_free) ; 

 %\IRule% [[gnu::pure]] Slot * ToSlot (uint64_t hapax) { 
 %\IRule%  %\IRule% static constexpr int ArraySize = 4096 ; 
 %\IRule%  %\IRule% static Slot WaitingArray alignas(4096) [ArraySize] {0} ; 
 %\IRule%  %\IRule% auto salt = uint32_t(uintptr_t(this)) ; 
 %\IRule%  %\IRule% uint32_t ix = ((salt + uint32_t(hapax >> 16)) * 17) & (ArraySize-1) ;
 %\IRule%  %\IRule% return WaitingArray + ix ; 
 %\IRule% }
 %\IRule%  
 %\IRule% std::atomic <uint64_t> Arrive {0} ;   // ingress -- most recent arrival
 %\IRule% std::atomic <uint64_t> Depart {0} ;   // egress -- most recent departure
 %\IRule%  
 %\IRule% static inline constinit thread_local uint64_t PrivateHapax = 0 ; 
 %\IRule% static inline constinit std::atomic<uint64_t> HapaxAllocator alignas(128) {0} ; 
 %\IRule%  
 public : 
 %\IRule% inline auto %{\fboxsep1pt\colorbox{yellow!50}{operator+}}%(std::invocable auto && csfn) %$\rightarrow$% void {
 %\IRule%  %\IRule%  
 %\IRule%  %\IRule% // Acquire the lock ...
 %\IRule%  %\IRule% // First, conjure a unique hapax identifier for this specific acquire-release episode
 %\IRule%  %\IRule% // The hapax is globally and temporally unique and specific to this thread, this lock, and this episode
 %\IRule%  %\IRule% %\HICC%auto hapax = PrivateHapax ++ ;                                                  %\TaggedT%
 %\IRule%  %\IRule% %\HICC%if ((hapax & 0xFFFF) == 0) [[unlikely]] {                                   %\TaggedT%
 %\IRule%  %\IRule%  %\IRule% // crossed edge of block allocation                                           
 %\IRule%  %\IRule%  %\IRule% // current block is exhausted so must reprovision                            
 %\IRule%  %\IRule%  %\IRule% // In this particular example the high 48-bits of the 64-bit hapax encode   
 %\IRule%  %\IRule%  %\IRule% // the thread "zone" and the lower 16 are the sub-sequence from which                     
 %\IRule%  %\IRule%  %\IRule% // the thread can allocate locally.  
 %\IRule%  %\IRule%  %\IRule% hapax = HapaxAllocator.fetch_add(1)+1 ; 
 %\IRule%  %\IRule%  %\IRule% // We should implement a proper overflow check and panic
 %\IRule%  %\IRule%  %\IRule% hapax = hapax << 16 ; 
 %\IRule%  %\IRule%  %\IRule% assert (hapax %$\neq$% 0) ; 
 %\IRule%  %\IRule%  %\IRule% assert (hapax > PrivateHapax) ;
 %\IRule%  %\IRule%  %\IRule% PrivateHapax = hapax + 1 ; 
 %\IRule%  %\IRule% }
 %\IRule%  %\IRule%  
 %\IRule%  %\IRule% assert (hapax %$\neq$% 0) ;       // by convention, 0 is reserved   
 %\IRule%  %\IRule% assert (hapax %$\neq$% Depart.load ()) ; 
 %\IRule%  %\IRule% assert (hapax %$\neq$% Arrive.load ()) ; 
 %\IRule%  %\IRule% %\HICC%auto pred = %\Global{Arrive}%.exchange (hapax) ;                                               %\TaggedT% 
 %\IRule%  %\IRule% assert (pred %$\neq$% hapax) ; 
 %\IRule%  %\IRule%  
 %\IRule%  %\IRule% %\HICC%if (%\Global{Depart}%.load(std::memory_order_acquire) %$\neq$% pred) {                               %\TaggedT%
 %\IRule%  %\IRule%  %\IRule% // Contention : slow-path waiting 
 %\IRule%  %\IRule%  %\IRule% // Initial lock acquire can not encounter contention ...
 %\IRule%  %\IRule%  %\IRule% // Pred is only 0 on the initial locking request
 %\IRule%  %\IRule%  %\IRule% // and we can't have contention on the 1st request, thus pred %$\neq$% 0 
 %\IRule%  %\IRule%  %\IRule% assert (pred %$\neq$% 0) ; 
 %\IRule%  %\IRule%  %\IRule% %\HICC%auto PredSlot = ToSlot(pred) ;                                                    %\TaggedT%
 %\IRule%  %\IRule%  %\IRule% %\HICC%if (PredSlot%$\rightarrow$%%\Global{VisibleWaiter}%.cas (0, pred) == 0) [[likely]] {                      %\TaggedT%
 %\IRule%  %\IRule%  %\IRule%  %\IRule% // successfully registered and emplaced ourselves as a visible waiter
 %\IRule%  %\IRule%  %\IRule%  %\IRule% // We now "own" and occupy the slot -- exclusively reserved 
 %\IRule%  %\IRule%  %\IRule%  %\IRule% // ratify -- Close race window vs thread in unlock() 
 %\IRule%  %\IRule%  %\IRule%  %\IRule% // We expect that Depart will remain, with high probability, resident
 %\IRule%  %\IRule%  %\IRule%  %\IRule% // in our local cache, given that we just fetched it, above.  
 %\IRule%  %\IRule%  %\IRule%  %\IRule% // Under contention, the 1st access, above, is liable to incur a coherence miss, 
 %\IRule%  %\IRule%  %\IRule%  %\IRule% // but the access below will usually be satisfied from the cache without a miss. 
 %\IRule%  %\IRule%  %\IRule%  %\IRule% %\HICC%if (%\Global{Depart}%.load() == pred) [[unlikely]]  {            %\TaggedT%
 %\IRule%  %\IRule%  %\IRule%  %\IRule%  %\IRule%  // Inopportune interleaving vs corresponding unlock() 
 %\IRule%  %\IRule%  %\IRule%  %\IRule%  %\IRule%  // Rescind visibility 
 %\IRule%  %\IRule%  %\IRule%  %\IRule%  %\IRule%  // Extricate ourselves from the Waiting array slot
 %\IRule%  %\IRule%  %\IRule%  %\IRule%  %\IRule%  // we raced vs unlock() and are now the owner -- release the slot
 %\IRule%  %\IRule%  %\IRule%  %\IRule%  %\IRule%  // We expect this to be rare 
 %\IRule%  %\IRule%  %\IRule%  %\IRule%  %\IRule%  // tempting to use store(0) but require cas(pred,0) 
 %\IRule%  %\IRule%  %\IRule%  %\IRule%  %\IRule%  // Need to use cas(pred,0) as racing unlock() might have already set value to 0
 %\IRule%  %\IRule%  %\IRule%  %\IRule%  %\IRule%  PredSlot%$\rightarrow$%%\Global{VisibleWaiter}%.cas (pred, 0) ; 
 %\IRule%  %\IRule%  %\IRule%  %\IRule% } else {
 %\IRule%  %\IRule%  %\IRule%  %\IRule%  %\IRule%  // Confirmed -- normal preferred mode of waiting
 %\IRule%  %\IRule%  %\IRule%  %\IRule%  %\IRule%  // The waiting thread is "settled" and emplaced
 %\IRule%  %\IRule%  %\IRule%  %\IRule%  %\IRule%  // Primary busy-wait waiting loop ... 
 %\IRule%  %\IRule%  %\IRule%  %\IRule%  %\IRule%  %\HICC%while (PredSlot%$\rightarrow$%%\Global{VisibleWaiter}%.load (std::memory_order_acquire) == pred) {   %\TaggedT% %$\blacktriangleleft$%
 %\IRule%  %\IRule%  %\IRule%  %\IRule%  %\IRule%  %\IRule%  %\HICC%Pause() ;                                                                  %\TaggedT%
 %\IRule%  %\IRule%  %\IRule%  %\IRule%  %\IRule%  %\HICC%}                                                                            %\TaggedT%
 %\IRule%  %\IRule%  %\IRule%  %\IRule% }
 %\IRule%  %\IRule%  %\IRule% } else {
 %\IRule%  %\IRule%  %\IRule%  %\IRule% // the slot is occupied by some other waiter because of a hash collision 
 %\IRule%  %\IRule%  %\IRule%  %\IRule% // We failed to install ourselves 
 %\IRule%  %\IRule%  %\IRule%  %\IRule% // we expect this situation to be rare
 %\IRule%  %\IRule%  %\IRule%  %\IRule% // Fallback and revert to either Tidex global spinning or HapaxLock waiting on slot%$\rightarrow$%useq
 %\IRule%  %\IRule%  %\IRule%  %\IRule% // Run as invisible waiter
 %\IRule%  %\IRule%  %\IRule%  %\IRule% while (%\Global{Depart}%.load (std::memory_order_acquire) %$\neq$% pred) {
 %\IRule%  %\IRule%  %\IRule%  %\IRule%  %\IRule% Pause()  ; 
 %\IRule%  %\IRule%  %\IRule%  %\IRule% }
 %\IRule%  %\IRule%  %\IRule% } 
 %\IRule%  %\IRule% } 
 %\IRule%  %\IRule%  
 %\IRule%  %\IRule% %{\rule[0pt]{0.65\linewidth}{.5pt}}% 
 %\IRule%  %\IRule% EnterCS: 
 %\IRule%  %\IRule% assert (Depart.load() %$\neq$% Arrive.load()) ;
 %\IRule%  %\IRule% // Execute the critical section, expressed as a lambda  
 %\IRule%  %\IRule% // Pass hapax value as context from lock to corresponding unlock
 %\IRule%  %\IRule% // In contention sustained steady-state, we expect just one thread
 %\IRule%  %\IRule% // to arrive during the CS 
 %\IRule%  %\IRule% // Arrival and depart rates should equilibrate
 %\IRule%  %\IRule% csfn() ;
 %\IRule%  %\IRule% assert (Depart.load() %$\neq$% Arrive.load()) ; 
 %\IRule%  %\IRule% %{\rule[0pt]{0.65\linewidth}{.5pt}}% 
 %\IRule%  %\IRule%  
 %\IRule%  %\IRule% // Unlock() ...
 %\IRule%  %\IRule% assert (Depart.load() %$\neq$% hapax) ; 
 %\IRule%  %\IRule%  
 %\IRule%  %\IRule% // Rendezvous with successor ...
 %\IRule%  %\IRule% // Attempt fast handover to a potential waiter -- successor
 %\IRule%  %\IRule% // inform waiting successor that we are renouncing or relinquishing ownership
 %\IRule%  %\IRule% // This is the preferred path for succession and handover under contention
 %\IRule%  %\IRule% // CONSIDER: hoist ToSlot(hapax) calculation up and before outside CS 
 %\IRule%  %\IRule% %\HICC%auto slot = ToSlot(hapax) ;                             %\TaggedT%
 %\IRule%  %\IRule% 
 %\IRule%  %\IRule% // CONSIDER: polite LD-if-ST vs CAS() 
 %\IRule%  %\IRule% %\HICC%if (slot%$\rightarrow$%%\Global{VisibleWaiter}%.cas (hapax, 0) == hapax) {      %\TaggedT% %$\blacktriangleleft$% 
 %\IRule%  %\IRule%  %\IRule% // success!                                           
 %\IRule%  %\IRule%  %\IRule% // We effected positive direct handover to our specific waiting successor 
 %\IRule%  %\IRule%  %\IRule% // We can safely elide the store into Depart()        
 %\IRule%  %\IRule%  %\IRule% %\HICC%return ;                                              %\TaggedT%
 %\IRule%  %\IRule% }
 %\IRule%  %\IRule%  
 %\IRule%  %\IRule% // Cases :
 %\IRule%  %\IRule% // @  no waiters -- uncontended case 
 %\IRule%  %\IRule% // @  waiter exists but we have hash collisions so our specific successor could 
 %\IRule%  %\IRule% //    not make itself visible
 %\IRule%  %\IRule% //    we expect this to be rare 
 %\IRule%  %\IRule% // @  waiter exists but our waiter was slow to make itself visible -- racy 
 %\IRule%  %\IRule% %\Global{Depart}%.store (hapax) ; 
 %\IRule%  %\IRule%  
 %\IRule%  %\IRule% // Need to CAS(hapax,0) on the wait array slot to close race window vs potential 
 %\IRule%  %\IRule% // tardy/sluggish arriving waiter.  
 %\IRule%  %\IRule% // Conservatively recover from potential race - compensate
 %\IRule%  %\IRule%  
 %\IRule%  %\IRule% // CONSIDER: ST-CAS vs polite ST-LD-if-CAS vs Exchange-if-CAS 
 %\IRule%  %\IRule% // If we use LD-if-CAS then either store() above or load() here must have full SC 
 %\IRule%  %\IRule% // std::memory_order_seq_cst fence semantics
 %\IRule%  %\IRule% // We need some type of store-load fence to avoid architectural races
 %\IRule%  %\IRule% // [store Depart; fence; load slot%$\rightarrow$%HapaxWaiter] 
 %\IRule%  %\IRule% // The complementary Dekker "pivot" executes in the wait arrival phase
 %\IRule%  %\IRule% slot%$\rightarrow$%%\Global{VisibleWaiter}%.cas (hapax, 0) ; 
 %\IRule% } 
} ;

\end{lstlisting}

Listing-\ref{Listing:Nonce-Visible-Waiters-FULL} reflects a variation that uses \emph{visible waiters}, 
which in turn enables assured \emph{positive handover}.  This particular expanded listing contains
augmented comments and optimizations.  

\newpage 

\section{Additional discussion regarding the \emph{Visible Waiters} variant} 

\Invisible{Dominant hot path}  

\Texticle{}Undesirable aspects of this variant, compared to baseline Hapax Locks, include the following :


\setlist[itemize,1]{label=$\bullet$} 
\begin{itemize} 

\item{} The visible waiters variant suffers from more complex paths, as there are races we need to recover from when 
threads start waiting at about the same time as their signaling unlock occurs. 

\item{} There are additional accesses to recheck shared globals, but 
those accesses tend to be to locations just accessed a few instructions ago and which are likely still 
resident in our thread's local cache.  
While our approach exhibits good spatial and temporal locality,
the benefits of that residency, in terms of coherence traffic, are timing-,
platform- and application-dependent.  

\Invisible{Coherence out-of-order ``transaction'' width; counted in accesss; arbitration window; group of accesses
covered by just once coherence transaction; coalesced coherence operations; atomicity granularity; 
Miss shadow; grace period; 
We posit and conjecture the existence of a small post coherence miss retention interval where the requestor can hold
the line against remote arbitration for a short period.  The existence is not guaranteed and is
likely an implementation artifact, but targetted ad-hoc experiments suggest it exists}  

\item{}The approach is relatively atomic-heavy and employs a number of CAS instructions.  
We note that counting atomics (CAS), once popular in concurrency literature -- and, at the time, relevant, given 
that legacy atomics were implemented via locking the bus --  may be somewhat anachronistic, given that modern processors 
(AMD Ryzen, apple M series) implement CAS locally, in cache, have much reduced the local latency and pipeline disruption induced 
by a CAS. Arguably, CAS isn't much different than a fenced store, and in particular the main 
cost of a CAS is, like a store, any necessary coherence-related write invalidation. 

\item{}Waiters completely occupy a slot for the duration of the waiting phase, which means we might be more 
exposed to collisions if we have a large number of waiters.   As such, slots can’t be shared amongst 
multiple concurrent waiters.   
We note, however, that there is almost no particular downside to increasing the size of the array to statistically
reduce the incidence of collisions.   
And we could, if desired, shift hapax zones if we detect collisions, in an agile and adaptive fashion,
to reduce the odds of future collisions, providing a form of collision avoidance.
Currently the code falls back to simple Tidex global spinning if a waiter finds the slot occupied, 
but we could be more refined and add an extra field to the slot, and just fall back to classic Hapax Locks (without 
visible waiters) in the event of collisions between waiters.  
So a slot would then have at most one ``privileged'' visible waiter but could support 
multiple invisible  waiters in the usual Hapax Lock fashion.

\item{} The uncontended \texttt{unlock} path is more complex, but in practice it doesn’t seem to make any performance 
difference as the path is just re-accessing locations that it already just touched.  

\end{itemize} 

\noindent{}Desirable aspects : 

\begin{itemize} 

\item{}We provide expedited handover under contention.  The critical paths in \texttt{unlock} 
and on the waiting side are short with few global accesses.  

\item{}\texttt{Unlock} can also skip updating the \texttt{Depart} (Egress) global if it can positively detect 
that it accomplished handover to a waiting successor.   The \emph{positive handover} property 
lets us safely skip the update of \texttt{Depart} in the \texttt{unlock} operation.  
This optimizations reduces write invalidations on the shared \texttt{Depart} variable.  
\texttt{unlock} only needs to update the \texttt{Depart} field when there is no contention or
if there were collisions in the waiting array.  

\end{itemize}

\Texticle{}In experiments with Hapax Locks with the number of threads set to more than the number of CPUs 
(intentionally forcing the system into preemption) and with the waiting array intentionally set to be small,
we find lots of preempted waiters occupying the waiting array slots which in turn forces other waiters 
to revert to the cruder global spinning waiting mode.  As noted above, however, performance
is dominated by the affects of preemption and lock design, and array collisions are not of particular importance. 
And, reiterating, the key tunable parameter for sizing the waiting array is the number of physical CPUs.

The only exception to that claim might be if the system is using a non-standard non-1:1 threading model, 
where, say, we had 1000 logical runnable virtual threads, but they were multiplexed over a smaller number 
of real kernel threads.  In that case we could probably end up saturating the waiting array and actually 
see some performance fall off because of array collisions and occupancy.   But at that point we would 
also been required to modify the locking subsystems to use virtual thread-aware parking mechanisms,
instead of spinning or traditional parking,
and the new parking mechanism would have be codesigned with the user-mode thread model and be able to 
voluntarily switch, in user mode, off the stalled virtual thread (waiting for the Hapax Lock) to some 
other runnable virtual thread.   
As such, Hapax Locks might not be an appropriate choice for environments that use non-1:1 user mode threading models.   

We observe, however, that Hapax Locks variant in Listing-\ref{Listing:Nonce-Visible-Waiters} could be 
made \emph{threading-model aware}, if, when waiting, immediately before they 
dismount from and abdicate a kernel thread (switching that kernel thread to carry another runnable
\footnote{A \emph{runnable} thread is ready and eligible for dispatch, but is not currently running on kernel thread}  
user-mode virtual) they also surrrender their slot in the waiting array in order to be altruistic and ``polite'' and decrease
the odds of collisions in the array for other running threads.  When the virtual thread is eventually dispatched and 
re-mounted on a kernel thread, it would then recheck \texttt{Depart}, and, if necessary, try to reclaim the slot
and reinstall itself in the array. 

\Invisible{Surrender; abdicate; yield; cede} 

\Texticle{}In order to reduce occupancy pressure in the global waiting array, we can optionally institute 
a 2-stage waiting strategy where an arriving thread will first busy-wait as a visible waiter, but after some period,
the thread will then voluntarily vacate its slot in the global array, to avoid monopolizing the element,
and switch to a longer-term waiting mode that uses invisible waiters, such as found in baseline Hapax locks.

That is, threads busy-wait briefly as visible waiters via the waiting array, but, if they have not been granted the lock
within a short period, they remove their visible waiting marker from the array, surrender the slot,
(retracting the visible waiting state indication) and revert to longer-term invisible waiting 
in a second global shared waiting array, which runs parallel to the primary visible waiting array.  
There, threads employ a second spinning phase, as invisible waiters, followed, if desired,
by \texttt{futex}-based waiting on the associated array element.    

The positive handover optimization enabled by visible waiters affords the most relative benefit 
under high flux situations, with high arrival and departure traffic rates on given lock. 
If threads fail to acquire the lock during the brief initial visible spinning phase, then
we know that throughput and progress over the contented lock is low, 
so the benefit of visible waiting and positive handover will not be as pronounced,
so reverting to invisible waiting does not appreciably degrade performance in this particular
operating regime.  But by shifting to invisible waiting, our threads can reduce load on the visible
waiting array, and allow other threads and locks to benefit from using positive handover via
the primary array.  

Each slot in the array could be provisioned with one field for exclusive visible waiting -- where the
visible waiter registers itself -- and another field configured in the fashion of baseline Hapax locks,
which allows multiple concurrent invisible waiters (arising from collisions) to monitor the slot. 
This allows the slot to be used in both modes.  
Alternatively, we could use 2 distinct arrays, running in parallel, one for visible waiting and another 
for invisible waiting. 
In addition, if a thread arrives and tries to register but finds the visible waiter
element occupied, it can fall back immediately to baseline invisible waiting, instead
of reverting to degenerate global spinning on \texttt{Depart}.  

As noted above, the visible waiting technique is preferred in the case of high flux locking, with
high arrival and traffic rates, as the fast positive handover property provided by 
visible waiting confers benefits for throughput over the contented lock, allowing
ownership to be conveyed more efficiently and with less latency.  
Using invisible waiters does not permit the fast positive, handover,
but if a thread has waited through its short-term grace period and switched from visible to invisible waiting,
it is likely that any reduction in performance from using invisible waiting is relatively negligible as
the lock is not enduring high throughput rates.  


\Invisible{phase change for lock from visible to invisible; emergent behavior}  
\Invisible{Altruism; Altruistic; polite; } 
\Invisible{Rescind; Retract; Cancel; Annul; Abandon; Abdicate; revert; vacate; cede; yield; surrender; withdraw; } 

\newpage 

\section{Expanded Hapax Lock listing -- with Invisible Waiters} 
\lstset{caption={Hapax Lock Variation with Invisible Waiters}}
\lstset{label={Listing:Nonce-InvisibleWaiters-FULL}}
\lstinputlisting[mathescape=true,escapechar=\%]{listing-NonceLock.cc-fx}  

Listing-\ref{Listing:Nonce-InvisibleWaiters-FULL} is an expanded listing augmented with additional comments and optimizations. 

\newpage

\end{document}